\documentclass[preprint,12pt]{elsarticle}

\usepackage{amssymb}
\usepackage{bbm}
\usepackage{mathrsfs}
\usepackage{pifont}
\usepackage{bbding}
\usepackage{mathrsfs}
\usepackage{stmaryrd}
\usepackage{amsmath}
\usepackage{amscd}
\usepackage[left,pagewise]{lineno}
\usepackage{graphicx}
\newtheorem{Th}{\bf Theorem}[section]
\newtheorem{Cor}[Th]{\bf Corollary}
\newtheorem{Def}[Th]{\bf Definition}
\newtheorem{Exam}[Th]{\bf Example}
\newtheorem{Rem}[Th]{\bf Remark}
\newtheorem{Lem}[Th]{\bf Lemma}
\newtheorem{Pro}[Th]{\bf Proposition}
\newcommand{\p}[1]{{\bf Proof.} #1 \ $\Box$}

\journal{}

\begin{document}

\begin{frontmatter}



\title{Fuzzy alternating $\mathrm{B\ddot{u}chi}$ automata over distributive lattices}

\author[label1]{Xiujuan Wei}
\ead{xiujuanwei@163.com}

\author[label1]{Yongming Li}
\ead{liyongm@snnu.edu.cn}

\address[label1]{College of Mathematics and Information Science, Shaanxi Normal University, Xi'an, Shaanxi 710119, PR China}

\begin{abstract}

We give a new version of fuzzy alternating $\mathrm{B\ddot{u}chi}$ automata over distributive lattices: weights are putting in every leaf node of run trees rather than along with edges from every node to its children. Such settings are great benefit to obtain complement just by taking dual operation and replacing each final weight with its complement. We prove that $L$-fuzzy nondeterministic $\mathrm{B\ddot{u}chi}$ automata have the same expressive power as $L$-fuzzy alternating $\mathrm{B\ddot{u}chi}$ ones. A direct construction (without related knowledge about $L$-fuzzy nondeterministic $\mathrm{B\ddot{u}chi}$ ones such as: above equivalence relation and their closure properties) is given to show that the languages recognized by $L$-fuzzy alternating co-$\mathrm{B\ddot{u}chi}$ automata are also $L$-fuzzy $\omega$-regular. Furthermore, the closure properties and the discussion about decision problems for fuzzy alternating $\mathrm{B\ddot{u}chi}$ automata are illustrated in our paper.

\end{abstract}


\end{frontmatter}


\section{Introduction}
\label{1}

In computation theory, nondeterminism has played important roles(\cite{10,13}). Viewing nondeterministic computations as words, systems and its specifications can be seen as languages, then we can translate problems about model checking, satisfiability and synthesis to ones about languages of automata. These transforms provide a new automata-theoretic approach to study system specification, verification and synthesis, and meanwhile such method is proven to be effective (\cite{22}). Nondeterministic computation has only existential quantifier, but as a generalization of nondeterminism,
``alternation", it has existential and universal quantifiers (\cite{3}). In \cite{3}, A.K.Chandra studied the properties about alternating Turing machines and their languages. Moreover, some information about alternating finite state automata and alternating pushdown automata were also introduced. Alternating automata is a useful model to study formal verification, and more information about it can be referred to \cite{12,20}.

In the study of linear temporal logic (\cite{22}), Vardi translated the problems about programs and specifications to the ones about languages of automata: he illustrated that alternating ($\mathrm{B\ddot{u}chi}$) automata have same expressive power as nondeterministic ($\mathrm{B\ddot{u}chi}$) ones, and the former ones are exponentially more succinct than the latters; The result automaton obtained after taking dual operation and exchanging final and non-final states is the complement to the original alternating ($\mathrm{B\ddot{u}chi}$) automaton, which reflects the great advantage of ``alternation". Then he use these conclusions to build an alternating $B\ddot{u}chi$ automaton for an LTL formula and such that the language of such automaton is exactly the set of computations satisfying that LTL formula.

Are these conclusions about alternating $\mathrm{B\ddot{u}chi}$ automata all suitable for weighted cases? i.e., (from the perspective of automata) Are there automata with weighted existential and universal quantifiers? In \cite{1,4}, O.Kupferman et al. had already introduced the definition of weighted alternating $\mathrm{B\ddot{u}chi}$ automata, which answers the above question. O.Kupferman et al. studied the expressive powers of weighted alternating $\mathrm{B\ddot{u}chi}$ automata for special semantics such as $Max$,
$Sum$,
$Sup$,
$LimSup$ and so on over real number set, and discussed the relationship between them simultaneously. But these specific semantics make the conclusions restricted, and the discussion about the relation between weighted alternating $\mathrm{B\ddot{u}chi}$ automata and nondeterministic $\mathrm{B\ddot{u}chi}$ ones is not involved. Furthermore, their automata have no final state, which is not comprehensive and general: the influences exerted by final states are not taken into the consideration, and thus, the Boolean cases cannot be seen as the special case of theirs. It shows the drawbacks of the version of weighted alternating $\mathrm{B\ddot{u}chi}$ automata in \cite{1,4}. So we want to give another one to avoid above shortcomings.

Derived from these ideas, we will introduce a new version of weighted alternating $\mathrm{B\ddot{u}chi}$ automata with weights in distributive lattices, of which the properties such as: the equivalence relation between weighted nondeterministic $\mathrm{B\ddot{u}chi}$ automata and weighted alternating $\mathrm{B\ddot{u}chi}$ ones, the closure properties can be established. In ours, the factor about final states are considered, and our version are more convenient to calculate the weights of their languages: for a word, to describe how likely it can be accepted depends on all successful runs on it, and the weight of each run is obtained just by taking conjunction of the weights of all branches, to be specific, if the branch is finite, its weight is equal to the label of its leaf node, otherwise, it is equal to $\bigwedge\limits_{i\geq 0}\bigvee\limits_{j\geq i}F(q_{j})$, where $q_{0},q_{1},\cdots$ is the label sequence of such branch and $F$ is the $L$-valued fuzzy sets of final states. Such advantage is due to our weights' and transitions' settings: the image set of the transition function ``$\delta$" is a subset of Boolean formulas over $L\cup Q$ ($L$ is a distributive lattice and $Q$ is the states set) rather than that in \cite{1,4}, a subset of Boolean formulas over $L\times Q$. Then in the runs of our version, weights and states are the labels of nodes (weights can and only can label the leaf nodes), which is much clearer and simpler than the case of \cite{1,4}: weights label the edges between nodes and each node is labeled by states.

In section 2, some pre-knowledge about alternating $\mathrm{B\ddot{u}chi}$ automata are introduced. In section 3, with the notion of fuzzy Boolean formulas, we give the definitions of fuzzy alternating $\mathrm{B\ddot{u}chi}$ automata over distributive lattices, show how to calculate the weights of run trees of our version (leaf nodes labeled by weights), and illustrate the equivalence relation between $L$-fuzzy alternating $\mathrm{B\ddot{u}chi}$ automata and $L$-fuzzy nondeterministic $\mathrm{B\ddot{u}chi}$ ones. The closure properties about $L$-fuzzy alternating $\mathrm{B\ddot{u}chi}$ automata are introduced in section 4. A construction showing the languages recognized by $L$-fuzzy alternating co-$\mathrm{B\ddot{u}chi}$ automata are also $L$-fuzzy $\omega$-regular without using the equivalence relation between $L$-fuzzy alternating $\mathrm{B\ddot{u}chi}$ automata and $L$-fuzzy nondeterministic $\mathrm{B\ddot{u}chi}$ ones and closure properties of $L$-fuzzy nondeterministic $\mathrm{B\ddot{u}chi}$ ones is provided. In section 5, we discuss the decision problems (emptiness-value, universality-value, implication-value problems) for $L$-fuzzy alternating $\mathrm{B\ddot{u}chi}$ automata: these problems can be decidable in exponential time and are PSPACE-complete. Some specific examples are given in the last section, which can be evidences to testify our theorems' correctness. Similarly to classical case, the above conclusions could also be seen as an effective approach to study fuzzy temporal logic, which can be leaving as one future study. For example, how to build a fuzzy alternating $\mathrm{B\ddot{u}chi}$ automaton for a fuzzy LTL formula such that the language of this automaton is exactly the fuzzy set of computations satisfying that fuzzy LTL formula.

\section{Preliminaries}
\label{2}

For a set $X$, let
$\mathbf{\mathcal{B}}^{+}(X)$ denote the set of all positive Boolean formulas over it (i.e., Boolean formulas built by elements of $X$ using $\wedge$ and $\vee$). Besides,
$\mathbf{\mathcal{B}}^{+}(X)$ includes two special formulas,
$\mathbf{true}$ and $\mathbf{false}$. For $Y\subseteq X$ and $\theta \in \mathbf{\mathcal{B}}^{+}(X)$, we say that $Y$ satisfies $\theta$, if the truth value is true after assigning $true$ to the members of $Y$ and assigning $false$ to the members of $X-Y$; furthermore, if there is no proper subset of $Y$ satisfying $\theta$, then we say $Y$ satisfies $\theta$ in a minimal manner. Obviously,
$\{x_{1},x_{2},x_{3}\}$ satisfies the formula $(x_{1}\vee x_{2})\wedge x_{3}$, and $\{x_{1},x_{2}\}$,
$\{x_{1},x_{3}\}$ satisfy it in a minimal manner, while the set $\{x_{2},x_{3}\}$ does not.

For any nondeterministic $\mathrm{B\ddot{u}chi}$ automaton $\mathcal{A}=(Q,\Sigma,\delta,q_{0},F)$, some formulas from $\mathbf{\mathcal{B}}^{+}(Q)$ can be used to represent its $\delta$. For example,
for a transition $\delta(q,a)=\{q_{1},q_{2},q_{3}\}$, it can be described by formula $q_{1}\vee q_{2}\vee q_{3}$. Based on such representation, there is a new notion: alternating $B\ddot{u}chi$ automata. The only distinctions between nondeterministic and alternating ones are transitions ``$\delta$".

\begin{Def} (\cite{22})
An alternating $\mathrm{B\ddot{u}chi}$ automaton is a five tuple $\mathcal{A}=(Q,\Sigma,\delta,q_{0},F)$, where $Q$ is a finite nonempty set of states,
$\Sigma$ is a finite nonempty set of input symbols, called alphabet;
$q_{0}$ and $F$ denote the initial state and the set of final states respectively,
$\delta$ is a transition function from $Q\times\Sigma$ into $\mathbf{\mathcal{B}}^{+}(Q)$.
\end{Def}

In an alternating $\mathrm{B\ddot{u}chi}$ automaton,
the transitions can be any formula of $\mathbf{\mathcal{B}}^{+}(Q)$. The language recognized by an alternating $\mathrm{B\ddot{u}chi}$ automaton is characterized by induction, for instance, if $\delta(q,a)=(q_{1}\wedge q_{2})\vee q_{3}$ is a transition of some alternating $\mathrm{B\ddot{u}chi}$ automaton, which means this automaton accepts $aw$ from $q$, if it accepts $w$ from both $q_{1}$ and $q_{2}$ or from $q_{3}$, where $w$ is a word of $\Sigma^{\omega}$. It is clear that such transition includes both the features of existential choice (the disjunction in the formula) and universal choice (the conjunction in the formula).

Because of the universal choice, a run of an alternating $\mathrm{B\ddot{u}chi}$ automaton is a tree rather than a sequence.
$|x|$ denotes the level which the node $x$ occurring at; in particular, for root $\varepsilon$,
$|\varepsilon|=0$ ($x$ and $\varepsilon$ are symbols rather than specific states). A branch $\beta=x_{0},x_{1},\cdots$ of a tree is a nodes sequence, where $x_{0}$ is $\varepsilon$ and $x_{i}$ is the parent of $x_{i+1}$ for all $i\geq 0$. In fact, a run $r$ of an alternating $\mathrm{B\ddot{u}chi}$ automaton is a $Q$-labeled tree, in which nodes are labeled by states.
$r(x)=q$ means that the node $x$ of $r$ labeled by $q$ ($x$ is a symbol and $q$ is a specific state).

\begin{Def}
A run of $\mathcal{A}$ on an infinite word $w=a_{0}a_{1}\cdots$ is a (possibly infinite) tree $r$ such that $r(\varepsilon)=q_{0}$ and the following holds:

If $|x|=i$,
$r(x)=q$, and $\delta(q,a_{i})=\theta$, then $x$ has $k$ children $x_{1},\cdots,x_{k}$,
for some $k\leq |Q|$, and $\{r(x_{1}),\cdots,r(x_{k})\}$ satisfies $\theta$ in a minimal manner.
\end{Def}

For example, if $\delta(q,a_{i})=(q_{1}\vee q_{2})\wedge q_{3}$, then the labels of $q$'s children include one element of $\{q_{1},q_{2}\}$ and also include state $q_{3}$ after putting $a_{i}$. Notice that if $\delta(r(x),a_{i})=\mathbf{true}$, then $x$ does not have any children, i.e.,
$x$ is a leaf node. In addition, there is no run taking a transition with $\theta=\mathbf{false}$.
The run tree $r$ is accepting if every infinite branch in $r$ infinitely passes $F$.

The relationships between alternating $\mathrm{B\ddot{u}chi}$ automata and nondeterministic $\mathrm{B\ddot{u}chi}$ automata have been studied (\cite{22}): they have the same expressive power, furthermore, the former ones are more succinct than latters, and the blow-ups of states during the transforms from alternating to nondeterministic ones are unavoidable (\cite{22}).

One advantage of alternating $\mathrm{B\ddot{u}chi}$ automata is that they are easy to be complemented. For equivalent alternating and nondeterministic $\mathrm{B\ddot{u}chi}$ automata, it is more easy to complement the former ones, cf.\cite{3}: just interchanging the conjunctions and disjunctions in every transition, as well as final and non-finial states.

\section{$L$-fuzzy alternating $\mathrm{B\ddot{u}chi}$ automata and their equivalent counterparts}
\label{3}

If not illustrate especially, the lattice $L$ we used below is distributive. In addition, we require that $L$ have the largest element $1$ and the least element $0$. In the following, we firstly introduce some preparation works. Our version of $L$-fuzzy alternating $\mathrm{B\ddot{u}chi}$ automata is distinct from \cite{1,4}: weights belong to $L$ rather than the real set, and in ours, weights labels every leaf node of run tree instead of along with every edge from each node to its child. In order to overcome shortcomings of \cite{1,4}, i.e., Boolean case cannot be seen as its special case, we put factor about final states in consideration.

\begin{Def}
An $L$-fuzzy positive Boolean formula over $X$ is a positive Boolean formula over $L\cup X$. The set of $L$-fuzzy positive Boolean formulas over $X$ is denoted by $\mathbf{\mathcal{F_{L}B}}^{+}(X)$, and moreover, we put the formulas $\mathbf{true}$ and $\mathbf{false}$ in it.
\end{Def}

For any $Y\subseteq X$ and a formula $\theta \in \mathbf{\mathcal{F_{L}B}}^{+}(X)$, we define a value $v(\theta,Y)$ in $L$, which is obtained by substituting any element of $Y$ occurring in $\theta$ by $1$, and that of $X-Y$ by $0$. Let $\theta_{1},\theta_{2}\in \mathbf{\mathcal{F_{L}B}}^{+}(X)$, if for any $Y\subseteq X$,
$v(\theta_{1},Y)=v(\theta_{2},Y)$ holds, then we call them equivalent, denoted by $\theta_{1}\equiv\theta_{2}$. For example, for
$\theta_{1}=0.5\vee(x_{2}\wedge 0.2\wedge x_{3})\vee(0.8\wedge x_{2})$ and
$\theta_{2}=0.5\vee(((0.3\wedge x_{3})\vee0.8))\wedge x_{2})$, we can verify that $\theta_{1}\equiv\theta_{2}$.

For any $\theta \in \mathbf{\mathcal{F_{L}B}}^{+}(X)$, it is easy to find its equivalent formula $\theta^{\prime}$, called standard form: in it each term between every two ``$\vee$" is in the form:
$l\wedge \bigwedge_{i\in I}x_{i}$ for some index set $I$ (if $l=1$, we always omit it and just write $\bigwedge_{i\in I}x_{i}$),
$l$ is a element in $L-\{0\}$, and we call it ``coefficient" of such term.

In fact, the factor impacting on the equivalence relation between formulas are their simplest final expansions: for above $\theta_{1}$ and $\theta_{2}$, they are equivalent because they have the identical simplest final expansions $0.5\vee (0.8\wedge x_{2})$. To be specific, we divide the procedures of obtaining the simplest final expansion for a given formula into the following steps:

$\mathbf{Step \ 1}$: Expand the formula;

$\mathbf{Step \ 2}$: Write above expansion in the standard form. In particular, there maybe exists a term $l$ ($l\in L$), called constant term, where its index set
$I=\emptyset$;

$\mathbf{Step \ 3}$: If there exist two term $l_{1}\wedge \bigwedge_{i\in I}x_{i}$ and $l_{2}\wedge \bigwedge_{j\in J}x_{j}$ such that $l_{1}\leq l_{2}$ and $J\subseteq I$, then remove the former one (indeed the former one is absorbed in the latters in the calculations of runs' weights).

Further on, we let $v(\mathbf{true},Y)=1$ for any set $Y$ (include empty set) and correspondingly, we let no set satisfy formula $\mathbf{false}$ (these settings are compatible with classic logic). Obviously,
$\mathbf{true}\equiv 1$.

For $\theta$, we define its satisfaction sets: if there exists a term $l\wedge \bigwedge_{i\in I}x_{i}$ in the standard form of $\theta$, we call $\{x_{i}|i\in I\}$ satisfies $\theta$ with weight $l$. Moreover, if it is also in the simplest final expansion of $\theta$, we say $\{x_{i}|i\in I\}$ satisfies $\theta$ in a minimal manner with weight $l$. In particular, for the constant term $l^{\prime}$ in the simplest final expansion, we call $\emptyset$ satisfies $\theta$ in a minimal manner with weight $l^{\prime}$. Also for formulas $\theta_{1}$ and $\theta_{2}$ mentioned above, we know $\{x_{2},x_{3}\}$ satisfies $\theta_{1}$ and $\theta_{2}$ with weights $0.2$ and $0.3$ respectively;
$\emptyset$ and $x_{2}$ satisfies $\theta_{1}$ and $\theta_{2}$ in a minimal manner with weights $0.5$ and $0.8$ respectively.

Considering an $L$-fuzzy nondeterministic $\mathrm{B\ddot{u}chi}$ automaton $\mathcal{A}=(Q,\Sigma,\delta,\\I,F)$, its transition function $\delta$ maps each state $q\in Q$ to an $L$-fuzzy set by inputting a symbol of $\Sigma$. We can represent $\delta$ by some formulas of $\mathbf{\mathcal{F_{L}B}}^{+}(Q)$: for example,
$\delta(q,a)=\frac{l_{1}}{q_{1}}+\frac{l_{2}}{q_{2}}+\frac{l_{3}}{q_{3}}$ (sometimes, we also use $\delta(q,a)(q_{i})=l_{i}$ or $\delta(q,a,q_{i})=l_{i}$ ($i=1,2,3$) to characterize such transition) can be described as $\delta(q,a)=(l_{1}\wedge q_{1})\vee (l_{2}\wedge q_{2})\vee (l_{3}\wedge q_{3})$ of $\mathbf{\mathcal{F_{L}B}}^{+}(Q)$. Generally, in an $L$-fuzzy alternating $\mathrm{B\ddot{u}chi}$ automaton,
the transitions can be any formula of $\mathbf{\mathcal{F_{L}B}}^{+}(Q)$.

\begin{Def}
An $L$-fuzzy alternating $\mathrm{B\ddot{u}chi}$ automaton is a tuple $\mathcal{A}=(Q,\Sigma,\delta,I,F)$, where $Q$ is a finite nonempty set of states,
$\Sigma$ is a finite nonempty alphabet,
$I$ and $F$ denote the $L$-valued fuzzy sets of initial and final states respectively, and
$\delta: Q\times \Sigma\rightarrow \mathbf{\mathcal{F_{L}B}}^{+}(Q)$ is an $L$-valued fuzzy transition function.
\end{Def}

\begin{Def}
A run of $\mathcal{A}$ on an infinite word $w=a_{0}a_{1}\cdots$ is a (possibly infinite) $(L\cup Q)$-labeled tree $r$ such that $I(r(\varepsilon))\neq 0$ and the following holds:

If $|x|=i$,
$r(x)=q$ and $\delta(q,a_{i})=\theta$, then $x$ has $k$ children $x_{1},\cdots,x_{k}$ for some $k\leq |Q|+1$ and $\{r(x_{1}),\cdots,r(x_{k})\}\cap Q$ satisfies $\theta$ in a minimal manner with weight $l\in\{r(x_{1}),\cdots,r(x_{k})\}\cap L$ (notice that the set $\{r(x_{1}),\cdots,r(x_{k})\}\cap L$ has at most one element, and if it is empty, this weight is $1$);

If $|x|=i$,
$r(x)=q$ and $\delta(q,a_{i})=\mathbf{true}$, then $x$ has one child $1$;

If $|x|=i$,
$r(x)=l$
($l\in L$), then the node $x$ has no children, i.e., it is a leaf (only nodes labeled by elements from $L$ can be leaves).
\end{Def}

For example, if $\delta(q,a)=(l_{1}\vee q_{2})\wedge q_{1}$, then $q$'s children are $l_{1}$ and $q_{1}$ or $q_{2}$ and $q_{1}$ after inputting $a$.

If the total weight of $r$ is not $0$, i.e.,
$weight(r)=I(r(\varepsilon))\wedge wt(r)\neq 0$, then we call $r$ an accepting run of $\mathcal{A}$, where $wt(r)$ is equal to the conjunction of all branches' weights in $r$. The weight of a branch $\beta$ is defined by:

If it is finite, its weight equals to $l$ ($\in L$), the label of the leaf node;

If it is infinite,
$\beta=x_{0},x_{1},\cdots$, and $r(x_{i})=q_{i}$, then $wt(\beta)$ equals to $\bigwedge\limits_{i\geq 0}\bigvee\limits_{j\geq i}F(q_{j})$.

Then for any $w\in\Sigma^{\omega}$,
$L_{\omega}(\mathcal{A})(w)=\bigvee\limits_{r\in R_{\mathcal{A}}(w)}I(r(\varepsilon))\wedge wt(r)$, where $R_{\mathcal{A}}(w)$ denotes the set of all runs on $w$ of $\mathcal{A}$.

\begin{Rem}
In $L$-fuzzy cases, we needn't require an alternating $\mathrm{B\ddot{u}chi}$ automaton to have a unique initial state, even though from the construction below, we know that every $L$-fuzzy alternating $B\ddot{u}chi$ automaton can be transformed to another equivalent one with a crisp initial state (which is sufficient for closure property in Section 4). In order to simulate $L$-fuzzy nondeterministic $B\ddot{u}chi$ automata, using Definition 3.2 is more accurately.
\end{Rem}

Here we give the corresponding construction (similar to \cite{18,19}):
let $\mathcal{A}=(Q,\Sigma,\delta,I,F)$ be an $L$-fuzzy alternating $\mathrm{B\ddot{u}chi}$ automaton, define an automaton with a crisp initial state $\mathcal{A}^{\prime}$ as $(Q\cup\{q_{0}\},\Sigma,\delta^{\prime},q_{0},F)$, where $q_{0}\notin Q$,
$\delta^{\prime}(q_{0},a)=\bigvee\limits_{I(q)\neq 0}I(q)\wedge \delta(q,a)$ and otherwise,
$\delta^{\prime}(q,a)=\delta(q,a)$.

\begin{Exam}
Let $\mathcal{A}=(Q,\Sigma,\delta,I,F)$ be an $L$-fuzzy alternating $\mathrm{B\ddot{u}chi}$ automaton, where $L=([0,1],\vee,\wedge,0,1)$;
$Q=\{q_{0},q_{1},q_{2},q_{3}\}$;
$\Sigma=\{a,b\}$;

$I(q_{0})=0.5$,
$I(q_{1})=I(q_{2})=I(q_{3})=0$;

$F(q_{0})=0$,
$F(q_{1})=0.4$,
$F(q_{2})=0.3$,
$F(q_{3})=0.1$;

$\delta(q_{0},a)=0.4\wedge q_{1}$,
$\delta(q_{0},b)=(0.5\wedge q_{2})\vee 0.3$,
$\delta(q_{1},a)=(0.2\wedge q_{1}\wedge q_{2})\vee (0.5\wedge q_{3})$;
$\delta(q_{1},b)=q_{2}$;
$\delta(q_{2},a)=0.2\wedge q_{1}\wedge q_{2}$,
$\delta(q_{2},b)=q_{3}$;
$\delta(q_{3},a)=q_{2}$,
$\delta(q_{3},b)=q_{3}$.

Set $w=a(ab)^{\omega}$. There are two successful run trees on $w$ and we denote them by $r,r^{\prime}$, then
$wt(r)$ and $wt(r^{\prime})$ are:

$wt(r)=\bigwedge\limits_{\beta \ is  \ finite\ in \ r}wt(\beta)\wedge\bigwedge\limits_{\beta \ is  \ infinite\ in \ r}wt(\beta)=0.2\wedge 0.3=0.2$;

$wt(r^{\prime})=\bigwedge\limits_{\beta \ is  \ finite\ in \ r}wt(\beta)\wedge\bigwedge\limits_{\beta \ is  \ infinite\ in \ r}wt(\beta)=0.3$.

\noindent{Hence,
$L_{\omega}(\mathcal{A})(w)=\bigvee\limits_{r\in R_{\mathcal{A}}(w)}I(r(\varepsilon))\wedge wt(r)=wt(r)\vee wt(r^{\prime})=0.3$.}

\begin{center}
  $\includegraphics[height=5.5cm,width=8cm]{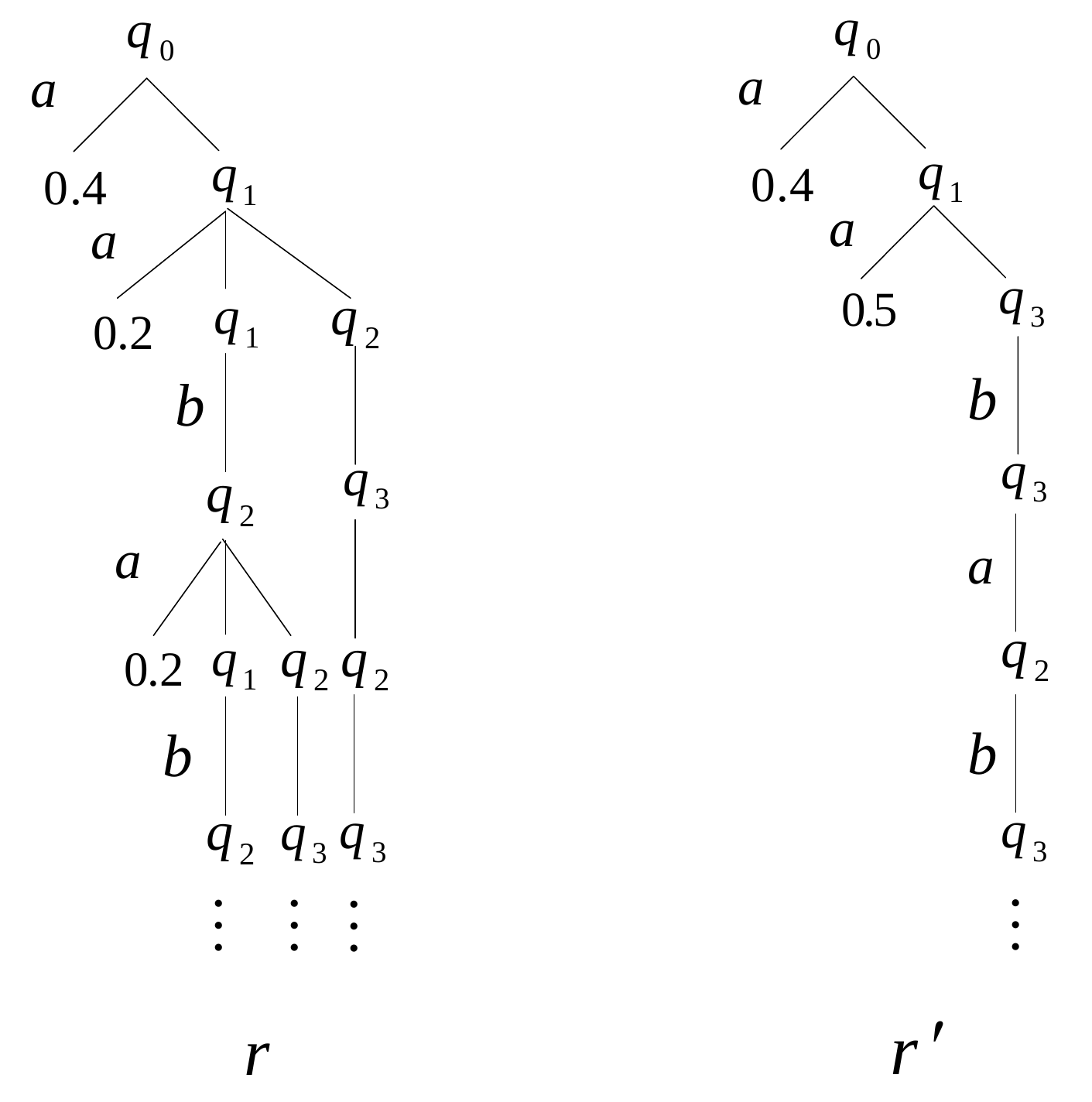}$
\end{center}
\begin{center}
  Figure 1: All successful runs of $\mathcal{A}$ on $a(ab)^{\omega}$
\end{center}

\end{Exam}

After introducing the basic definitions, we are ready to study the equivalence relation between fuzzy nondeterministic $\mathrm{B\ddot{u}chi}$ automata and fuzzy alternating $\mathrm{B\ddot{u}chi}$ automata over distributive lattices. Firstly, we show that $L$-fuzzy alternating $\mathrm{B\ddot{u}chi}$ automata are at least as expressive and as succinct as $L$-fuzzy nondeterministic $\mathrm{B\ddot{u}chi}$ automata.

\begin{Pro}
Assume that $\mathcal{A}$ is an $L$-fuzzy nondeterministic $B\ddot{u}chi$ automaton with $n$ states, then there is an $L$-fuzzy alternating $B\ddot{u}chi$ automaton $\mathcal{A}_{a}$ with $n$ states such that $L_{\omega}(\mathcal{A}_{a})=L_{\omega}(\mathcal{A})$.
\end{Pro}

\p{Let $\mathcal{A}=(Q,\Sigma,\delta,I,F)$ be the given $L$-fuzzy nondeterministic $\mathrm{B\ddot{u}chi}$ automaton. Define an $L$-fuzzy alternating $\mathrm{B\ddot{u}chi}$ automaton $\mathcal{A}_{a}=(Q,\Sigma,\delta_{a},I,F)$:
where $\delta_{a}(q,b)=\bigvee\limits_{\delta(q,b)(q^{\prime})=l_{q^{\prime}}\neq 0}l_{q^{\prime}}\wedge q^{\prime}$,
$b\in \Sigma$, and otherwise, if $\delta(q,b)(q^{\prime})=0$ for any $q^{\prime}\in Q$, we set $\delta_{a}(q,b)=\mathbf{false}$.

Let $w$ be an arbitrary word of $\Sigma^{\omega}$ (denoted by $w=a_{1}a_{2}\cdots$) such that $L(\mathcal{A})(w)\neq 0$. Assume that $P$ is a run on $w$ of $\mathcal{A}$ such that $weight(P)\neq 0$, i.e., there a sequence of states $q,q_{1},q_{2},\cdots$ such that
$I(q)\neq 0$,
$\delta(q,a_{1},q_{1})\neq 0$ and $\delta(q_{i},a_{i+1},q_{i+1})\neq 0$ ($i\geq1$),
$\bigwedge\limits_{i\geq0}\bigvee\limits_{j\geq i}F(q_{j})\neq 0$, then there exists a corresponding successful run tree $r$ on $w$ of $\mathcal{A}_{a}$ satisfying:

At $0$-th level of $r$, there is only one element $q$, and $I(r(\varepsilon))=I(q)\neq 0$;

At $1$-th level of $r$, there are two elements: a leaf node labeled by $\delta(q,a_{1},q_{1})$ and a non-leaf node labeled by $q_{1}$;

$\cdots$

At $i$-th level of $r$, there are two elements: a leaf node labeled by $\delta(q_{i-1},a_{i},q_{i})$ and a non-leaf node labeled by $q_{i}$;

$\cdots$.

Then we have,

\begin{eqnarray*}
& &I(r(\varepsilon))\wedge wt(r)\\
&=&I(r(\varepsilon))\wedge \delta(r(\varepsilon),a_{1},q_{1})\wedge \bigwedge\limits_{i\geq1} \delta(q_{i},a_{i},q_{i+1})\wedge \bigwedge\limits_{i\geq0} \bigvee\limits_{j\geq i} F(q_{j})\\
&=&I(q)\wedge \delta(q,a_{1},q_{1})\wedge \bigwedge\limits_{i\geq1} \delta(q_{i},a_{i},q_{i+1})\wedge \bigwedge\limits_{i\geq0} \bigvee\limits_{j\geq i} F(q_{j}),
\end{eqnarray*}

\noindent{and thus $L(\mathcal{A})(w)=L(\mathcal{A}_{a})(w)$.}

Conversely, we can also show that $L(\mathcal{A})(w)=L(\mathcal{A}_{a})(w)$, for any $w\in\Sigma^{\omega}$ such that $L(\mathcal{A}_{a})(w)\neq 0$.}

This part is easy to be obtained, and afterwards, we will turn to the other one. We divides it into two steps: firstly, we shall prove that any $L$-fuzzy alternating $\mathrm{B\ddot{u}chi}$ automaton with crisp final states can be transformed to an equivalent $L$-fuzzy nondeterministic $\mathrm{B\ddot{u}chi}$ automaton; Secondly, we will show that every $L$-fuzzy alternating $\mathrm{B\ddot{u}chi}$ automaton can be converted to another one with crisp final states. The next proposition shows the first step:

\begin{Pro}
For any $L$-fuzzy alternating $B\ddot{u}chi$ automaton $\mathcal{A}$ with $n$ states, if it has crisp final states, then there is an $L$-fuzzy nondeterministic $B\ddot{u}chi$ automaton $\mathcal{A}_{n}$ with at most $3^{n}$ states satisfying $L_{\omega}(\mathcal{A}_{n})=L_{\omega}(\mathcal{A})$.
\end{Pro}

\p{Let $\mathcal{A}=(Q,\Sigma,\delta,I,F)$ be an $L$-fuzzy alternating $\mathrm{B\ddot{u}chi}$ automaton, where $F$ is a crisp set of  final states. Define $\mathcal{A}_{n}=(Q_{n},\Sigma,\delta_{n},I_{n},F_{n})$ as follows:
$Q_{n}=2^{Q}\times 2^{Q}$;
for any $q\in Q$, we let $I_{n}((\{q\},\emptyset))=I(q)$, and otherwise,
$I_{n}((A,B))=0$, where $A,B\in 2^{Q}$;
$F_{n}=2^{Q}\times \{\emptyset\}$;

For any $(U,V)\in Q_{n}$,
$V\neq \emptyset$, and if $U=\{q_{1},\cdots,q_{s}\}$,
$V=\{q^{\prime}_{1},\cdots,q^{\prime}_{m}\} (\subseteq U)$, we define $\delta_{n}$ by:

\noindent{\begin{eqnarray*}
& &\delta_{n}((U,V),a,(U^{\prime},V^{\prime}))\\
&=&(\bigvee\limits_{\substack{\{q_{i_{1}}^{\prime},\cdots,
q_{i_{l_{i}}}^{\prime}\}\subseteq U^{\prime};\\
i=1,\cdots,m}} \bigwedge\limits_{i=1}^{m}\mu_{l_{i}}(a)_{q_{i_{1}}^{\prime}\cdots q_{i_{l_{i}}}^{\prime},q_{i}^{\prime}}))\wedge (\bigvee\limits_{\substack{\{q_{j_{1}},\cdots,
q_{j_{l_{j}}}\}\subseteq U^{\prime};\\
q_{j}\in U-V}} \bigwedge\limits_{q_{j}\in U-V}\mu_{l_{j}}(a)_{q_{j_{1}}\cdots q_{j_{l_{j}}},q_{j}}),
\end{eqnarray*}}

\noindent{where $\bigcup\limits_{i=1}^{m}\{q_{i_{1}}^{\prime},\cdots,
q_{i_{l_{i}}}^{\prime}\}-F=V^{\prime}$,
$\bigcup\limits_{j=1}^{s}\{q_{j_{1}},\cdots,
q_{j_{l_{j}}}\}=U^{\prime}$, and $U^{\prime}$ is a set satisfying the conjunction of all simplest final expansions of all $\delta(t,a)(t\in U)$,
$X$ is a set satisfying the conjunction of all simplest final expansions of all $\delta(t,a)(t\in V)$,
$V^{\prime}=X-F$.}

For any $(U,\emptyset)\in Q_{n}\times Q_{n}$, if $U=\{p_{1},\cdots,p_{k}\}$,
then $\delta_{n}$ is defined as:

\begin{eqnarray*}
\delta_{n}((U,\emptyset),a,(U^{\prime},V^{\prime}))=\bigvee\limits_{\substack{\{p_{i_{1}},\cdots,p_{i_{l_{i}}}\}
\subseteq U^{\prime};\\i=1,\cdots,k}}\bigwedge_{i=1}^{k}\mu_{l_{i}}(a)_{p_{i_{1}},\cdots,p_{i_{l_{i}}},p_{i}},
\end{eqnarray*}

\noindent{where $\bigcup\limits_{i=1}^{s}\{p_{i_{1}},\cdots,p_{i_{l_{i}}}\}=U^{\prime}$,
$U^{\prime}-F=V^{\prime}$, and $U^{\prime}$ is a set satisfying conjunction of all simplest final expansions of all $\delta(t,a)(t\in U)$.}

We take an empty conjunction in the definition of $\delta_{n}$ to be $1$, i.e.,
$\delta_{n}((\emptyset,\emptyset),a,\\(\emptyset,\emptyset))=1$. In addition, the others not mentioned are defined to $0$.

On one hand, we need to prove that for any $w\in \Sigma^{\omega}$, if it satisfies $L_{\omega}(\mathcal{A}_{n})(w)\neq 0$, then $L_{\omega}(\mathcal{A}_{n})(w)=L_{\omega}(\mathcal{A})(w)$.

In fact, for any successful run $P:(\{q\},\emptyset) \stackrel{a_{1}}{\rightarrow}(A_{1},B_{1})\stackrel{a_{2}}{\rightarrow}(A_{2},B_{2})\rightarrow\cdots$, we can construct a run $r$ of $\mathcal{A}$:

Put $r(\varepsilon)=q$ firstly;

Let all states of $A_{1}$ ($A_{1}\triangleq\{q_{11},\cdots,q_{1s}\}$) be the children of $q$ occurring at $1$-th level of $r$;

If $B_{1}\triangleq\{\widehat{q}_{11},\cdots,\widehat{q}_{1m}\}\neq \emptyset$, we follow the steps below:

Let $\delta(q_{1i},a_{2})=\theta_{1i}$,
$\delta(\widehat{q}_{1j},a_{2})=\widehat{\theta}_{1j}$,
$i=1,\cdots,s$,
$j=1,\cdots,m$, we choose sets $B_{2j}=\{\widehat{q}_{j1}^{\prime},\cdots,\widehat{q}_{jl_{j}}^{\prime}\}\subseteq A_{2}$ such that
$B_{2j}$ satisfies $\widehat{\theta}_{1j}$ in a minimal manner with weight $\mu_{l_{j}}(a_{2})_{\widehat{q}_{j1}^{\prime}\cdots \widehat{q}_{jl_{j}}^{\prime},\widehat{q}_{1j}}$,
$j=1,\cdots,m$, and $\bigcup\limits_{j=1}^{m}B_{2j}-F=B_{2}$. Meanwhile, we choose sets $A_{2i}=\{q_{i1}^{\prime},\cdots,q_{il_{i}}^{\prime}\}\subseteq A_{2}$ such that
$A_{2i}$ satisfies $\theta_{1i}$ in a minimal manner with weight $\mu_{l_{i}}(a_{2})_{q_{i1}^{\prime}\cdots q_{il_{i}}^{\prime},q_{1i}}$,
$i=1,\cdots,s$,
$\bigcup\limits_{i=1}^{s}A_{2i}=A_{2}$ and if there are $t_{1},t_{2}$ such that $q_{1t_{1}}=\widehat{q}_{1t_{2}}$, then $A_{2t_{1}}=B_{2t_{2}}$. Then we let $\mu_{l_{i}}(a_{2})_{q_{i1}^{\prime}\cdots q_{il_{i}}^{\prime},q_{1i}},q_{i1}^{\prime},\cdots,q_{il_{i}}^{\prime}$ be the children of $q_{1i}$ occurring at $2$-th level of $r$.

If $B_{1}=\emptyset$, we just choosing sets $A_{2i}=\{q_{i1}^{\prime},\cdots,q_{il_{i}}^{\prime}\}\subseteq A_{2}$ such that
$A_{2i}$ satisfies $\theta_{1i}$ in a minimal manner with weight $\mu_{l_{i}}(a_{2})_{q_{i1}^{\prime}\cdots q_{il_{i}}^{\prime},q_{1i}}$,
$i=1,\cdots,s$, and
$\bigcup\limits_{i=1}^{s}A_{2i}=A_{2}$,
$\bigcup\limits_{i=1}^{s}A_{2i}-F=B_{2}$. Then we let $\mu_{l_{i}}(a_{2})_{q_{i1}^{\prime}\cdots q_{il_{i}}^{\prime},q_{1i}},q_{i1}^{\prime},\cdots,q_{il_{i}}^{\prime}$ be the children of $q_{1i}$ occurring at $2$-th level of $r$.

Similarly, the choices of other levels are considered.

We observe that even though the run tree constructed is not unique (under isomorphism), the disjunction of all these probabilities' total weights is equal to
$weight(P)=I_{n}((\{q\},\emptyset))\wedge \delta((\{q\},\emptyset),a_{1},(A_{1},\cdots,B_{1}))\wedge \bigwedge\limits_{i\geq 1}\delta((A_{i},B_{i}),a_{i+1},\\(A_{i+1},B_{i+1}))$ (because $L$ is distributive). Then we have:

\begin{eqnarray*}
L_{\omega}(\mathcal{A})(w)&=&\bigvee_{r\in R_{\mathcal{A}}(w)}\bigwedge_{\beta\ is \ a \ branch\ of \ r}wt(\beta)\\
&=&\bigvee_{P\in R_{\mathcal{A}_{n}}(w)}(\bigvee_{r\in R(P)} \bigwedge_{\beta\ is \ a \ branch\ of \ r} wt(\beta))
\end{eqnarray*}

\begin{eqnarray*}
&=&\bigvee_{P\in R_{\mathcal{A}_{n}}(w)}(\bigvee_{r\in R(P)}wt(r))\\
&=&\bigvee_{P\in R_{\mathcal{A}_{n}}(w)}weight(P)\\
&=&L_{\omega}(\mathcal{A}_{n})(w),
\end{eqnarray*}

\noindent{where $R_{\mathcal{A}}(w)$ and $R_{\mathcal{A}_{n}}(w)$ denote the set of all runs on $w$ of $\mathcal{A}$ and $\mathcal{A}_{n}$ respectively, and $R(P)$ denotes the set of all runs of $\mathcal{A}$ constructed by $P$.}

On the other hand, for any successful run $r$ of $\mathcal{A}$ on a infinite word $w=a_{1}a_{2}\cdots$, we can construct a run $P^{\prime}$ of $\mathcal{A}_{n}$:

$\widehat{A_{0}}=(r(\varepsilon),\emptyset)$;

$\widehat{A_{1}}=(A_{1},B_{1})$ (where $A_{1}=\{q| q \ is \ the \ child \ of \ r(\varepsilon)\}$,
$B_{1}= \{q| q \ is  \ the\\ \ child \ of \ r(\varepsilon)\}-F$);

If $B_{1}\neq \emptyset$, we let $\widehat{A_{2}}=(A_{2},B_{2})$ (where $A_{2}=\{q| q \ is \ the \ child \ of \ some \ state\\ \ of  \ A_{1}\}$,
$B_{2}=\{q| q \ is \ the \ child \  of \ some \ state \ of \ B_{1}\}-F$), and otherwise, we set $\widehat{A_{2}}=(A_{2},A_{2}-F)$;

$\cdots$

Similarly, there may be several run trees corresponding to such $P^{\prime}$ of $\mathcal{A}_{n}$, but the disjunction of their total weights is equal to $weight(P^{\prime})$, then we have $L_{\omega}(\mathcal{A}_{n})(w)=L_{\omega}(\mathcal{A})(w)$ likewise.

Obviously, we can find that for each reachable state $(U,V)$ of $\mathcal{A}_{n}$, then $V\subseteq U$, and thus the number of states in $\mathcal{A}_{n}$ is at most $3^{n}$.}

Notice that in above proof, in ``$U^{\prime}$ is a set satisfying the conjunction of all simplest final expansions of all $\delta(t,a)(t\in U)$,
$X$ is a set satisfying the conjunction of all simplest final expansions of all $\delta(t,a)(t\in V)$", such ``satisfying" needn't be required ``in a minimal manner", in fact, if we add such requirement, it may loss some non-zero possibilities of transitions. For example, suppose that $U=\{q_{1},q_{2}\}$, $V=\emptyset$, and $q_{1},q_{2},q_{3}$ are final states, if the simplest final expansion of $\delta(q_{1},a)$ is $(q_{1}\wedge q_{3})\vee(0.3\wedge q_{2}\wedge q_{3})$, and that of $\delta(q_{2},a)$ is $(0.1\wedge q_{1})\vee(0.2\wedge q_{2})$, then $\delta_{n}((\{q_{1},q_{2}\},\emptyset),a,(\{q_{1},q_{2},q_{3}\},\emptyset))=0.2$ according to Proposition 3.7. If we add requirement ``in a minimal manner", we will obtain that $\delta((\{q_{1},q_{2}\},\emptyset),a,(\{q_{1},q_{2},q_{3}\},\emptyset))=0$, which destroys the equivalence relation that we want to obtain.

The first goal has been reached, then the last question need to be resolved is that: how to transform an ordinary $L$-fuzzy alternating $\mathrm{B\ddot{u}chi}$ automaton to another one with crisp final states.

\begin{Lem}
Let $\mathcal{A}_{1}$ and $\mathcal{A}_{2}$ be $L$-fuzzy alternating $B\ddot{u}chi$ automata with crisp final states over $\Sigma$ and they have $n_{1}$ and $n_{2}$ states respectively, then there is another $L$-fuzzy alternating $B\ddot{u}chi$ automaton over $\Sigma$ with $n_{1}+n_{2}$ states,
$\mathcal{A}_{\vee}$, such that it also has crisp final states and satisfies $L_{\omega}(\mathcal{A}_{\vee})=L_{\omega}(\mathcal{A}_{1})\vee L_{\omega}(\mathcal{A}_{2})$.
\end{Lem}

\p{Let $\mathcal{A}_{1}=(Q_{1},\Sigma,\delta_{1},I_{1},F_{1})$ and $\mathcal{A}_{2}=(Q_{2},\Sigma,\delta_{2},I_{2},F_{2})$, where $F_{1}$ and $F_{2}$ are crisp final sets. Without loss of generality, we assume that $Q_{1}\cap Q_{2}=\emptyset$. Then
$\mathcal{A}_{\vee}$ is defined as $(Q_{1}\cup Q_{2},\Sigma,\delta,I,F_{1}\cup F_{2})$, where $\delta(q,a)=\delta_{i}(q,a)$, if $q\in Q_{i}$ for some $i$;
$I(q)=I_{i}(q)$, if $q\in Q_{i}$ for some $i$.

For any $w\in\Sigma^{\omega}$, we can prove that $L_{\omega}(\mathcal{A}_{\vee})(w)=L_{\omega}(\mathcal{A}_{1})(w)\vee L_{\omega}(\mathcal{A}_{2})(w)$. In fact, for any successful run of $\mathcal{A}_{\vee}$, then it is also a successful one of $\mathcal{A}_{i}$ for some $i$, and conversely, all successful runs of $\mathcal{A}_{1}$ and $\mathcal{A}_{2}$ are also successful in $\mathcal{A}_{\vee}$.}

\begin{Pro}
Suppose that $\mathcal{A}$ is an $L$-fuzzy alternating $B\ddot{u}chi$ automaton with $n$ states, then there is an equivalent $L$-fuzzy alternating $B\ddot{u}chi$ automaton $\mathcal{A^{\prime}}$ with $n\cdot \sum\limits_{i=0}^{k}C_{n}^{i}$ states such that $\mathcal{A^{\prime}}$ has crisp final states, where $k=|supp(F)-ker(F)|$,
$F$ is the fuzzy final states set of $\mathcal{A}$, and $ker(F)=\{q|F(q)=1\}$.
\end{Pro}

\p{According to Remark 3.4, we only need to focus our attention on any fuzzy alternating $\mathrm{B\ddot{u}chi}$ automaton with a crisp initial state.

Assume that $\mathcal{A}=(Q,\Sigma,\delta,q_{0},F)$, where $|supp(F)-ker(F)|=k$. For any $s\leq k$, we define a set $s(Q)$, which contains all choices of different $s$ states from $supp(F)-ker(F)$ and all members of $ker(F)$, i.e., $s(Q)=\{\{q_{i_{1}},\cdots,q_{i_{s}}\}\cup ker(F)|q_{i_{1}},\cdots,q_{i_{s}}\in supp(F)-ker(F), and \ q_{t}\neq q_{t^{\prime}}\ if \ t \neq t^{\prime}\}$.

For any $s\leq k$, any element $P\in s(Q)$ (denoted by $\{q_{j_{1}},\cdots,q_{j_{s}}\}\cup ker(F)$), we define an $L$-fuzzy alternating $\mathrm{B\ddot{u}chi}$ automaton with crisp final states $\mathcal{A}_{P}=(Q,\Sigma,\delta,I_{P},F_{P})$:

$I_{P}(q_{0})=\bigwedge\limits_{i=1}^{s}F(q_{j_{i}})$ and otherwise,
$I_{P}(q)=0$;
$F_{P}=P=\{q_{j_{1}},\cdots,q_{j_{s}}\}\cup ker(F)$.

In the following, we point out $L_{\omega}(\mathcal{A})=\bigvee\limits_{P\in ker(F)\cup1(Q)\cup\cdots\cup k(Q)}L_{\omega}(\mathcal{A}_{p})$. Let $r$ be an infinite run tree of $\mathcal{A}$, we know $r$ is also a run tree of each $\mathcal{A}_{P}$.

If

\begin{eqnarray*}
weight(r)&=&\bigwedge\limits_{\beta\ is \ a\ branch \ of \ r}wt(\beta)\\
&=&\bigwedge\limits_{\substack{\beta\ is \ a\ branch \ of \ r,\\ and\ \beta\  finitely\ passes\
ker(F)}}wt(\beta)\\
&=& l_{1}\wedge\cdots\wedge l_{t}\wedge (\bigvee\limits_{i_{1}\in I_{1}}F(q_{i_{1}}))\wedge\cdots\wedge (\bigvee\limits_{i_{l}\in I_{l}}F(q_{i_{l}}))\\
&=&\bigvee\limits_{i_{1}\in I_{1},\cdots,i_{l}\in I_{l}}l_{1}\wedge\cdots\wedge l_{t}\wedge F(q_{i_{1}})\wedge\cdots\wedge F(q_{i_{l}}),
\end{eqnarray*}

\noindent{then there exist $l$ branches of $r$,
$\beta_{1},\cdots,\beta_{l}$, such that $\beta_{1}$ infinitely passes $q_{i_{1}}$ (for any $i_{1}\in I_{1}$),
$\cdots$,
$\beta_{l}$ infinitely passes $q_{i_{l}}$ (for any $i_{l}\in I_{l}$).
Therefore,

\begin{eqnarray*}
\bigvee_{P\in P_{l}}weight_{\mathcal{A}_{P}}(r)=weight(r),
\end{eqnarray*}

And for any $P$ of $ker(F)\cup1(Q)\cup\cdots\cup k(Q)-P_{l}$, we have:

\begin{eqnarray*}
weight_{\mathcal{A}_{P}}(r)\leq weight(r),
\end{eqnarray*}

\noindent{where $P_{l}=\{\{q_{i_{1}^{\prime}},\cdots,q_{i_{l}^{\prime}}\}\cup ker(F)|i_{1}^{\prime}\in I_{1},\cdots,i_{l}^{\prime}\in I_{l}\}$ (there may exist $t_{1}\neq t_{2}$ such that $q_{i_{t_{1}}^{\prime}}=q_{i_{t_{2}}^{\prime}}$, if so, $|\{q_{i_{1}^{\prime}},\cdots,q_{i_{l}^{\prime}}\}|<\ min \{l,k\}$).}} Above all, we obtain

\begin{eqnarray*}
weight(r)=\bigvee\limits_{P\in ker(F)\cup1(Q)\cup\cdots\cup k(Q)}weight_{\mathcal{A}_{P}}(r).
\end{eqnarray*}

Even though for any $r$, there is a $P_{l}$ corresponding to it, on the whole, the parameter $P_{l}$ has no effect on value $weight(r)=\bigvee\limits_{P\in ker(F)\cup1(Q)\cup\cdots\cup k(Q)}weight_{\mathcal{A}_{P}}(r)$.

The above $s$ may be $0$, if so,
$I_{ker(F)}(q_{0})=1$ and $F_{ker(F)}=ker(F)$, then only runs whose all branches infinitely pass $ker(F)$ are successful infinite runs of $\mathcal{A}_{ker(F)}$.

If $weight(r)=1$, i.e., all branches infinitely pass $ker(F)$, and at this time, we have $weight_{\mathcal{A}_{P}}(r)=weight(r)=1$ for any $P\in ker(F)\cup1(Q)\cup\cdots\cup k(Q)$, therefore, the following equation also holds:

\begin{eqnarray*}
weight(r)=\bigvee\limits_{P\in ker(F)\cup1(Q)\cup\cdots\cup k(Q)}weight_{\mathcal{A}_{P}}(r).
\end{eqnarray*}

According to the definition of run, we know that there may be finite-depth runs on some $w\in \Sigma^{\omega}$. And in this case, every finite-depth run $r$ of $\mathcal{A}$ is also a successful run tree of each $\mathcal{A}_{P}$ (including $\mathcal{A}_{ker(F)}$), and $weight_{\mathcal{A}_{ker(F)}}(r)=weight(r)$ holds; for any $P\neq ker(F)$,
$weight_{\mathcal{A}_{P}}(r)\leq weight(r)$. Then we also have $weight(r)=\bigvee\limits_{P\in ker(F)\cup1(Q)\cup\cdots\cup k(Q)}weight_{\mathcal{A}_{P}}(r)$ for the finite-depth case.

Hence for any $w\in \Sigma^{\omega}$, we obtain:

\begin{eqnarray*}
L_{\omega}(\mathcal{A})(w)&=&\bigvee\limits_{r\in R_{\mathcal{A}}(w)}weight(r)\\
&=&\bigvee\limits_{r\in R_{\mathcal{A}}(w)}(\bigvee\limits_{P\in ker(F)\cup1(Q)\cup\cdots\cup k(Q)}weight_{\mathcal{A}_{P}}(r))\\
&=&\bigvee\limits_{P\in ker(F)\cup1(Q)\cup\cdots\cup k(Q)}(\bigvee\limits_{r\in R_{\mathcal{A}(w)}}weight_{\mathcal{A}_{P}}(r))\\
&=&\bigvee\limits_{P\in ker(F)\cup1(Q)\cup\cdots\cup k(Q)}(\bigvee\limits_{r\in R_{\mathcal{A}_{P}(w)}}weight_{\mathcal{A}_{P}}(r))\\
&=&\bigvee\limits_{P\in ker(F)\cup1(Q)\cup\cdots\cup k(Q)}L_{\omega}(\mathcal{A}_{P})(w).
\end{eqnarray*}

Set $\mathcal{A}^{\prime}=\bigvee\limits_{P\in ker(F)\cup1(Q)\cup\cdots\cup k(Q)}\mathcal{A}_{P}$, then we know that such $\mathcal{A}^{\prime}$ is our desired fuzzy alternating $\mathrm{B\ddot{u}chi}$ automaton according to Lemma 3.8.}

Putting Proposition 3.7 and 3.9 together, we have:

\begin{Th}
Assume that $\mathcal{A}$ is an $L$-fuzzy alternating $B\ddot{u}chi$ automaton with $n$ states, then there is an equivalent $L$-fuzzy nondeterministic $\mathrm{B\ddot{u}chi}$ automaton $\mathcal{A}^{\prime}$ with at most $3^{n\cdot \sum\limits_{i=0}^{k}C_{n}^{i}}$ states, where $k=|supp(F)-ker(F)|$,
$F$ is the fuzzy final states set of $\mathcal{A}$ and $ker(F)=\{q|F(q)=1\}$.
\end{Th}

\section{Closure properties of $L$-fuzzy alternating $\mathrm{B\ddot{u}chi}$ automata}
\label{4}

In this section, we study closure properties of $L$-fuzzy alternating $\mathrm{B\ddot{u}chi}$ automata. We show that $L$-fuzzy alternating $\mathrm{B\ddot{u}chi}$ automata are closed under join, meet and complementation. Firstly, we discuss the first two operations.

\begin{Th}
Let $\mathcal{A}_{1}$ and $\mathcal{A}_{2}$ be $L$-fuzzy alternating $B\ddot{u}chi$ automata over $\Sigma$, with $n_{1}$ and $n_{2}$ states, respectively. There are two $L$-fuzzy alternating $\mathrm{B\ddot{u}chi}$ automata $\mathcal{A}_{\vee}$ and $\mathcal{A}_{\wedge}$ over $\Sigma$, with $n_{1}+n_{2}$ and $n_{1}+n_{2}+1$
states respectively, such that $L_{\omega}(\mathcal{A}_{\vee})=L_{\omega}(\mathcal{A}_{1})\vee L_{\omega}(\mathcal{A}_{2})$ and $L_{\omega}(\mathcal{A}_{\wedge})=L_{\omega}(\mathcal{A}_{1})\wedge L_{\omega}(\mathcal{A}_{2})$.
\end{Th}

\p{According to Remark 3.4 and Proposition 3.9, it's enough to discuss the ones with one crisp initial state and crisp final states. Let $\mathcal{A}_{i}=(Q_{i},\Sigma,\delta_{i},(q_{0})^{(i)},F_{i})$. Without loss of generality, we assume that these two $Q_{i}$ are disjointed. Define $\mathcal{A}_{\vee}=(Q_{1}\cup Q_{2},\Sigma,\delta,\{(q_{0})^{(1)},(q_{0})^{(2)}\},F_{1}\cup F_{2})$:
$\delta(q,a)=\delta_{i}(q,a)$, for any $q\in Q_{i}$ and $a\in \Sigma$. Obviously, the following proof is analogous to that in Lemma 3.8, and we omit it here.

Let $\mathcal{A}_{\wedge}=(Q_{1}\cup Q_{2}\cup\{q_{0}\},\Sigma,\delta^{\prime},q_{0},F_{1}\cup F_{2})$ , of which $q_{0}\notin Q_{1}\cup Q_{2}$ and $\delta$ is defined as:
$\delta(q_{0},a)=\delta_{1}((q_{0})^{(1)},a)\wedge \delta_{2}((q_{0})^{(2)},a)$ and $\delta(q,a)=\delta_{i}(q,a)$, for any $q\in Q_{i}$ and $a\in \Sigma$.

Then $L_{\omega}(\mathcal{A}_{\wedge})(w)=L_{\omega}(\mathcal{A}_{1})(w)\wedge L_{\omega}(\mathcal{A}_{2})(w)$ can be got easily for any $w\in \Sigma^{\omega}$.}

As we all know, one advantage of alternating ($\mathrm{B\ddot{u}chi}$) automata is that it is easy to complement them. Is this advantage also suitable for $L$-fuzzy case? Indeed, we can demonstrate the dual of an $L$-fuzzy alternating $\mathrm{B\ddot{u}chi}$ automaton, an $L$-fuzzy alternating co-$\mathrm{B\ddot{u}chi}$ automaton, recognizes the complement of the language of the original automaton by game-theory. Notice that the following lattice has a negation $c$, which is a mapping from $L$ to $L$, satisfying $l_{1}\leq l_{2}\Rightarrow c(l_{2})\leq c(l_{1})$ and $c(c(l))=l$, for any $l,l_{1},l_{2}\in L$. The complement of fuzzy language $L(\mathcal{A})$, denoted by $L(\mathcal{A})^{c}$, is defined as $L(\mathcal{A})^{c}(w)=c(L(\mathcal{A})(w))$, for any $w\in \Sigma^{\omega}$.

Also because of Remark 3.4, we only need to focus on any $L$-fuzzy alternating $\mathrm{B\ddot{u}chi}$ automaton with a crisp initial state in following.

Let $\mathcal{A}=(Q,\Sigma,\delta,q_{0},F)$ be a such one, we define its dual, an $L$-fuzzy alternating co-$\mathrm{B\ddot{u}chi}$ automaton, denoted by $\overline{\mathcal{A}}$, where $\overline{\mathcal{A}}=(Q,\Sigma,\overline{\delta},q_{0},F^{c})$,
and $\overline{\delta}(q,a)=\overline{\delta(q,a)}$ for all $q\in Q$,
$a\in\Sigma$. Moreover,
$F^{c}(q)=c(F(q))$, for any $q\in Q$, where $c$ is the negation of $L$. The dual operation $\overline{\delta}$ is defined as:

$-\overline{q}=q$, for $q\in Q$;

$-\overline{l}=c(l)$, for any $l\in L$ (in particular,
$\overline{1}=0$ and $\overline{0}=1$);

$-\overline{(\alpha\wedge\beta)}=(\overline{\alpha}\vee\overline{\beta})$ and

$-\overline{(\alpha\vee\beta)}=(\overline{\alpha}\wedge\overline{\beta})$;

$-\overline{\mathbf{true}}=\mathbf{false}$;

$-\overline{\mathbf{false}}=\mathbf{true}$.

Let $\mathcal{B}$ be an $L$-fuzzy alternating co-$\mathrm{B\ddot{u}chi}$ automaton
$(Q,\Sigma,\delta,q_{0},F)$, the definition of runs is identical to that of $\mathrm{B\ddot{u}chi}$ one, but the successful runs and the calculation of their weights are different:

If the total weight of $r$ is not $0$, i.e.,
$weight(r)=I(r(\varepsilon))\wedge wt(r)\neq 0$, then we call $r$ an accepting run of $\mathcal{B}$, where $wt(r)$ equals to the conjunction of all branches' weights. The weight of a branch $\beta$ is defined by:

If it is finite,
$wt(\beta)$ equals to $l$ ($\in L$), the label of the leaf node;

If it is infinite,
$\beta=x_{0},x_{1},\cdots$, and $r(x_{i})=q_{i}$, then its weight equals to $\bigvee\limits_{i\geq 0}\bigwedge\limits_{j\geq i}F(q_{j})$.

Then for any $w\in\Sigma^{\omega}$,
$L_{\omega}(\mathcal{B})(w)=\bigvee\limits_{r\in R_{\mathcal{B}}(w)}I(r(\varepsilon))\wedge wt(r)$, where $R_{\mathcal{B}}(w)$ denotes the set of all runs on $w$ of $\mathcal{B}$.

Notice that the acceptance condition of $\overline{\mathcal{A}}$ is a fuzzy co-$\mathrm{B\ddot{u}chi}$ acceptance condition rather than fuzzy $\mathrm{B\ddot{u}chi}$ acceptance condition of $\mathcal{A}$, then for an infinite branch $\beta=x_{0},x_{1},\cdots$, and $r(x_{i})=q_{i}$, its weight is (we use subscripts to distinguish the weights of $\mathcal{A}$ and $\overline{\mathcal{A}}$):

\begin{eqnarray*}
wt_{\overline{\mathcal{A}}}(\beta)=\bigvee\limits_{i\geq 0}\bigwedge\limits_{j\geq i}F^{c}(q_{j})=c(\bigwedge\limits_{i\geq 0}\bigvee\limits_{j\geq i}F(q_{j}))=c(wt_{\mathcal{A}}(\beta)).
\end{eqnarray*}

Indeed,
the language recognized by such $L$-fuzzy alternating co-$\mathrm{B\ddot{u}chi}$ automaton is also $L$-fuzzy $\omega$-regular, i.e., it can also be recognized by an $L$-fuzzy alternating $\mathrm{B\ddot{u}chi}$ automaton. We will show it after Theorem 4.2.

\begin{Th}
Let $\mathcal{A}$ be an $L$-fuzzy alternating $B\ddot{u}chi$ automaton, then \\ $L_{\omega}(\overline{\mathcal{A}})(w)=c(L_{\omega}(\mathcal{A})(w))$ for any $w\in\Sigma^{\omega}$.
\end{Th}

\p{Similarly, we only consider the one with a crisp initial state. Let $\mathcal{A}=(Q,\Sigma,\delta,q_{0},F)$ be such an automaton. The value of a word $w$ ($w=a_{1}a_{2}\cdots$) in $\mathcal{A}$ can be thought as the outcome of following two-players (Player OR and Player AND) game. The game starts from initial state $q_{0}$ of $\mathcal{A}$. In every round, Player OR chooses a set $E\subseteq Q$ satisfying $\delta(q_{i},a_{i})$ in a minimal manner with weight $l$. Player AND chooses a state $q_{i+1}\in E$, and the game goes on from $q_{i+1}$ likewise. The goal of Player OR is to ``maximize" the value (corresponds to the supremum in different runs), and the goal of Player AND is to ``minimize" it (corresponds to the infimum in different branch of a run). The branch induced by this game corresponds to a ``minimal" branch (infimum) in a supreme run of $\mathcal{A}$.

When the same game is played on $\overline{\mathcal{A}}$, these two players interchange their actions' orders. The branch induced by this game corresponds to a ``maximal" branch in a ``minimal" run trees of $\mathcal{A}$ on $w$. Indeed, the Player AND determines which branch is taken in every run tree of $\mathcal{A}$ firstly, and Player OR determines which run is taken afterwards. Because of the fact that ``$wt_{\overline{\mathcal{A}}}(\beta)=\bigvee\limits_{i\geq 0}\bigwedge\limits_{j\geq i}F^{c}(q_{j})=c(\bigwedge\limits_{i\geq 0}\bigvee\limits_{j\geq i}F(q_{j}))=c(wt_{\mathcal{A}}(\beta))$" mentioned before, we know that the weight of every branch $c(l)$ in $\overline{\mathcal{A}}$ corresponds to the one,
$l$, in $\mathcal{A}$. Then, for every word $w\in \Sigma^{\omega}$, we have:

\begin{eqnarray*}
L_{\omega}(\overline{\mathcal{A}})(w)&=&\bigwedge\limits_{r\in R_{\mathcal{A}}(w)}\bigvee\limits_{\beta \ is \ a \ branch\ of \ r}wt_{\overline{\mathcal{A}}}(\beta)\\
&=&\bigwedge\limits_{r\in R_{\mathcal{A}}(w)}\bigvee\limits_{\beta \ is \ a \ branch\ of \ r}c(wt_{\mathcal{A}}(\beta))\\
&=&\bigwedge\limits_{r\in R_{\mathcal{A}}(w)}(c(\bigwedge\limits_{\beta \ is \ a \ branch\ of \ r}wt_{\mathcal{A}}(\beta)))\\
&=&c(\bigvee\limits_{r\in R_{\mathcal{A}}(w)}\bigwedge\limits_{\beta \ is \ a \ branch\ of \ r}wt_{\mathcal{A}}(\beta))\\
&=&c(L_{\omega}(\mathcal{A})(w)).
\end{eqnarray*}

Note that the first equation is obtained by above discussion rather than the definition. It shows the relationship between the same game playing in $\mathcal{A}$ and $\overline{\mathcal{A}}$.}

According to the statements ``all fuzzy languages recognized by $L$-fuzzy nondeterministic $\mathrm{B\ddot{u}chi}$ automata over $\Sigma$ is closed under complement" (Theorem 12 of \cite{11}) and ``equivalence relationship between $L$-fuzzy alternating $\mathrm{B\ddot{u}chi}$ automata and $L$-fuzzy nondeterministic $\mathrm{B\ddot{u}chi}$ ones" (c.f. Proposition 3.6 and Theorem 3.10), we obtain that the language recognized by an $L$-fuzzy alternating co-$\mathrm{B\ddot{u}chi}$ automaton is also recognized by an $L$-fuzzy alternating $\mathrm{B\ddot{u}chi}$ automaton, which shows the closure property about complement of $L$-fuzzy alternating $\mathrm{B\ddot{u}chi}$ automata.

Go a step further, we can give the direct construction to illustrate the $L$-fuzzy $\omega$-regularity of the languages recognized by $L$-fuzzy alternating co-$\mathrm{B\ddot{u}chi}$ automata without the knowledge about $L$-fuzzy nondeterministic $\mathrm{B\ddot{u}chi}$ automata.

\begin{Lem}
Every $L$-fuzzy alternating co-$B\ddot{u}chi$ automaton can be converted to another equivalent one with a crisp initial state.
\end{Lem}

\begin{Cor}
Let $\mathcal{A}$ be an $L$-fuzzy alternating co-$B\ddot{u}chi$ automaton, then $L_{\omega}(\overline{\mathcal{A}})(w)=c(L_{\omega}(\mathcal{A})(w))$ for any $w\in\Sigma^{\omega}$.
\end{Cor}

The proofs of Lemma 4.3 and Corollary 4.4 are similar to Remark 3.4 and Theorem 4.2 respectively, so we omit them here.

Taking twice dual operations and taking complement on final weights, we can get the following proposition, which is a co-$\mathrm{B\ddot{u}chi}$ version of Proposition 3.9, and it is the first step of our construction.

\begin{Pro}
Let $\mathcal{A}$ be an $L$-fuzzy alternating co-$B\ddot{u}chi$ automaton with $n$ states. Then there is an equivalent $L$-fuzzy alternating co-$B\ddot{u}chi$ automaton $\mathcal{A^{\prime}}$ with $1+n\cdot \sum\limits_{i=0}^{k}C_{n}^{i}$ states such that it has a crisp initial state and crisp final states (where $k=|supp(F)-ker(F)|$,
$F$ is the fuzzy final states set of $\overline{\mathcal{A}}$, and $ker(F)=\{q|F(q)=1\}$.).
\end{Pro}

To be specific, the procedures to get $\mathcal{A^{\prime}}$ are:

$\mathbf{Step \ 1}$: Following Corollary 4.4, we construct the dual of $\mathcal{A}$, an $L$-fuzzy alternating $\mathrm{B\ddot{u}chi}$ automaton, and it satisfies $L_{\omega}(\overline{\mathcal{A}})=c(L_{\omega}(\mathcal{A}))$;

$\mathbf{Step \ 2}$: Following Proposition 3.9, we construct an equivalent $L$-fuzzy alternating $\mathrm{B\ddot{u}chi}$ automaton $\mathcal{B}$ with crisp final states;

$\mathbf{Step \ 3}$: Following Remark 3.4, we construct an $L$-fuzzy alternating $\mathrm{B\ddot{u}chi}$ automaton $\mathcal{B}^{\prime}$ with a crisp initial state and crisp final states such that $L_{\omega}(\mathcal{B}^{\prime})=L_{\omega}(\mathcal{B})$;

$\mathbf{Step \ 4}$: Following Theorem 4.2, we construct the dual of $\mathcal{B}^{\prime}$, an $L$-fuzzy alternating co-$\mathrm{B\ddot{u}chi}$ automaton with crisp final states, and it satisfies that $L_{\omega}(\overline{\mathcal{B}^{\prime}})=c(L_{\omega}(\mathcal{B}^{\prime}))$.

Let $\mathcal{A^{\prime}}=\overline{\mathcal{B}^{\prime}}$, and it is our desired automaton.

Before giving Theorem 4.10, we need to introduce some notions firstly (cf. \cite{12}). In order to let the following content be compatible with front sections of our paper, we set the definitions and pre-knowledge version below are a bit different from \cite{12}, mainly reflecting on the final states set $F$. More detailed information can be referred to \cite{12}.

Let $\mathcal{A}$ be an alternating co-$\mathrm{B\ddot{u}chi}$ automaton. For nodes $x_{1}$ and $x_{2}$ of an accepting run $r$ of $\mathcal{A}$, we call that $x_{1}$ and $x_{2}$ are similar if and only if $|x_{1}|=|x_{2}|$ and $r(x_{1})=r(x_{2})$. Furthermore, $r$ is called memoryless if and only if the subtrees rooted at nodes $x_{1}$ and $x_{2}$ are identical for all similar nodes $x_{1}$ and $x_{2}$ of $r$.

\begin{Pro}
For an $L$-fuzzy co-$B\ddot{u}chi$ automaton $\mathcal{A}$, if there is a successful run $r$ on $w$ of it with total weight $l$, then there exists a memoryless accepting one on $w$ with total weight $l^{\prime}$ which is larger than or equal to $l$.
\end{Pro}

Proposition 4.6 tells us only memoryless accepting ones have effect on value $L_{\omega}(\mathcal{A})(w)$ (the others are absorbed in memoryless ones), then in the following, we only consider the memoryless runs.

Replacing label $q$ of node $x_{i}$ by $\langle q,i\rangle$ where $i=|x_{i}|$, and merging similar nodes into a single one, then we get an directed acyclic graph $G_{r}$ with respect to a memoryless run $r$. In $r$, if there is a states sequence $q_{0}$,
$q_{1}$,
$q_{2}$,
$\cdots$ (partial labels of nodes of some branch in $r$) such that $q=q_{0}$ and
$q^{\prime}=q_{i}$ ($i\geq 0$), then we say that $\langle q^{\prime},l^{\prime}\rangle$ is reachable from $\langle q,l\rangle$ in $G_{r}$.

Considering a directed acyclic graph $G\subseteq G_{r}$, a vertex $\langle q,l\rangle$ is said to be endangered in $G$ if and only if finitely many vertices in $G$ are reachable from $\langle q,l\rangle$;
$\langle q,l\rangle$ is safe in $G$ if and only if the projections of all the vertices that are reachable from $\langle q,l\rangle$ in $G$ on $Q$ belong to $F$ (the final states of $\mathcal{A}$). With these notions, we define an sequence of directed acyclic graphs $G_{0}\supseteq G_{1}\supseteq G_{2}\supseteq \cdots$ inductively as follows:

$-G_{0}=G_{r}$;

$-G_{2i+1}=G_{2i}\setminus \{\langle q,l\rangle|\langle q,l\rangle\ is \ endangered \ in \ G_{2i} \}$

$-G_{2i+2}=G_{2i+1}\setminus \{\langle q,l\rangle|\langle q,l\rangle\ is \ safe \ in \ G_{2i+1} \}$.

From \cite{12}, we know that for any vertex $\langle q,l\rangle$ in $G_{r}$, there is a unique index $i\geq 1$ such that $\langle q,l\rangle$ is either endangered in $G_{2i}$ or safe in $G_{2i+1}$. Then for each vertex, there is a notion ``rank", which describe such $i$:

$rank(\langle q,l\rangle)=\left \{  \begin {array}
             {r@{\quad \quad}l}
2i,& if\ \langle q,l\rangle \ is \ endangered \ in\ G_{2i}.\\
2i+1,& if\ \langle q,l\rangle \ is \ safe \ in\ G_{2i+1}.\\
\end{array} \right. \nonumber$

The two lemmas below show us the close connection between the ranks and the reachability of vertices, and they are the key to our last theorem.

\begin{Lem}(\cite{12})
For every two vertices $\langle q,l\rangle$ and $\langle q^{\prime},l^{\prime}\rangle$ in $G_{r}$, if $\langle q^{\prime},l^{\prime}\rangle$ is reachable from $\langle q,l\rangle$, then $rank(\langle q,l\rangle)\leq rank(\langle q^{\prime},l^{\prime}\rangle)$.
\end{Lem}

\begin{Lem}(\cite{12})
In every infinite path of $G_{r}$, there exists a vertex $\langle q,l\rangle$ with an odd rank such that all the vertices $\langle q^{\prime},l^{\prime}\rangle$ in the path that are reachable from $\langle q,l\rangle$ have $rank(\langle q^{\prime},l^{\prime}\rangle)=rank(\langle q,l\rangle)$.
\end{Lem}

At last, we will give the construction to get an equivalent $L$-fuzzy (weak) alternating $\mathrm{B\ddot{u}chi}$ automaton from an $L$-fuzzy alternating co-$\mathrm{B\ddot{u}chi}$ one.

\begin{Def}
An $L$-fuzzy weak alternating $B\ddot{u}chi$ automaton is a five tuple $(Q,\Sigma,\delta,I,F)$, where the states set $Q$ is some disjoint unions,
$Q_{i}$ ($i\in I$), and on these $Q_{i}$ there is a partial order $\leq$; in addition,
the transition function $\delta$ satisfies that: if $q\in Q_{i}$ and $q^{\prime}$ occurs in $\delta(q,a)$, then $q^{\prime} \in Q_{j}$ and $Q_{i}\leq Q_{j}$;
$F$ is $L$-fuzzy function from $Q_{i}$ to $L$.
\end{Def}

Note that the $L$-fuzzy weak alternating $\mathrm{B\ddot{u}chi}$ automaton is a special $L$-fuzzy alternating $\mathrm{B\ddot{u}chi}$ automaton, and its specificity reflects on states space, which is divided into several disjointed partially ordered sets. Moreover, it requires that every $q$ goes to the state which is in a smaller set than that $q$ stays in.

\begin{Th}
Let $\mathcal{A}$ be an $L$-fuzzy alternating co-$B\ddot{u}chi$ automaton with $n$ states, then there is an $L$-fuzzy weak alternating $B\ddot{u}chi$ automaton $\mathcal{A}^{\prime}$ with $2n^{2}$ states such that $L_{\omega}(\mathcal{A}^{\prime})=L_{\omega}(\mathcal{A})$.
\end{Th}

\p{From Proposition 4.5, we just consider some one with a crisp initial state and crisp final states. Let $\mathcal{A}=(Q,\Sigma,\delta,q_{0},F)$ be a such one, where $|Q|=n$. Define an $L$-fuzzy weak alternating $\mathrm{B\ddot{u}chi}$ automaton $\mathcal{A}^{\prime}=(Q^{\prime},\Sigma,\delta^{\prime},q_{0}^{\prime},F^{\prime})$:
$Q^{\prime}=Q\times [2n]$ ($[2n]=\{1,\cdots,2n\}$),
$q_{0}^{\prime}=(q_{0},2n)$,
$F^{\prime}=Q\times [2n]^{odd}$ ($[2n]^{odd}=\{1,3,\cdots,2n-1\}$).

The transition function $\delta^{\prime}$ is described by a function ``$release$", which is a mapping from $\mathbf{\mathcal{F_{L}B}}^{+}(Q)\times [2n]$ to $\mathbf{\mathcal{F_{L}B}}^{+}(Q^{\prime})$: for any $\theta\in \mathbf{\mathcal{F_{L}B}}^{+}(Q)$, a rank $i\in [2n]$, the formula $release(\theta,i)$ is obtained by replacing every $q$ in $\theta$ by $\bigvee\limits_{i^{\prime}\leq i}(q,i^{\prime})$, and then $\delta^{\prime}$ is defined by:

$\delta((q,i),a)=\left \{  \begin {array}
             {r@{\quad \quad}l}
release(\delta(q,a),i),& if\ q\in F \ or \  i \ is \ even.\\
false,& if\ q \notin F \ and \ i \ is \ odd.\\
\end{array} \right. \nonumber$

For each rank $i$, we put
$Q_{i}=Q\times \{i\}$. Obviously, for every state $(q,i)\in Q^{\prime}$, its possible children only belong to $L\cup Q_{i^{\prime}}$ ($i^{\prime}\leq i$).

Next, we shall prove $L_{\omega}(\mathcal{A})(w)=L_{\omega}(\mathcal{A})(w)$ for any $w\in \Sigma^{\omega}$.

Indeed, for any $w=a_{1}a_{2}\cdots\in \Sigma^{\omega}$ such that $L_{\omega}(\mathcal{A})(w)\neq 0$, there is at least a successful run of $\mathcal{A}$, denoted by $r$ (where $r(\varepsilon)=q_{0}$). Define a run of $\mathcal{A}^{\prime}$ as follows:

Let $r^{\prime}(\varepsilon)=(q_{0},2n)$;

If the children of $r(\varepsilon)$ are $\mu_{q_{1}\cdots q_{k},r(\varepsilon)},q_{1},\cdots,q_{k}$, i.e.,
$\{q_{1},\cdots,q_{k}\}$ satisfies $\delta(r(\varepsilon),a_{1})$ in a minimal manner with weight $\mu_{k}(a_{1})_{q_{1}\cdots q_{k},r(\varepsilon)}$, then we let $\mu_{k}(a_{1})_{q_{1}\cdots q_{k},r(\varepsilon)},(q_{1},i_{1}),\cdots,(q_{k},i_{k})$ be children of $r^{\prime}(\varepsilon)$ at $1$-th level of $r^{\prime}$, and $i_{j}$ be any one less than or equal to $2n$,
$j=1,\cdots,k$. Similarly, the choices of the states at other levels follow the same way. Note that the definition of ranks ensures
``$r(x) \notin F$" and ``$rank(\langle r(x),|x|\rangle)$ is odd" cannot hold simultaneously.

Among the runs constructed by above procedures, there is at least a successful one (all these successful runs' weights are equal to $weight(r)$). In fact, a run $r^{\ast}$ is a such one, in which the label of $x_{i}$ is $( r(x_{i}),rank(\langle r(x_{i}),|x_{i}|\rangle))$,
$i\geq1$. Lemma 4.7 and 4.8 ensure that $r^{\ast}$ is successful, and thus we obtain $L_{\omega}(\mathcal{A})(w)=L_{\omega}(\mathcal{A}^{\prime})(w)$.

Conversely, it just need to consider the projection of any successful run $r^{\prime}$ of $\mathcal{A}^{\prime}$ on $Q$, and $L_{\omega}(\mathcal{A})(w)=L_{\omega}(\mathcal{A}^{\prime})(w)$ holds similarly. This part is easy to show, and we omit it.}

\section{Decision problems for $L$-fuzzy alternating $\mathrm{B\ddot{u}chi}$ automata}
\label{5}

The aim of this section is to discuss decision problems for $L$-fuzzy alternating $\mathrm{B\ddot{u}chi}$ automata over a distributive lattice with a negation $c$. These discussions can be applied to the satisfiability and model-checking problem of fuzzy LTL \cite{15,16}.

Considering an $L$-fuzzy alternating $\mathrm{B\ddot{u}chi}$ automaton $\mathcal{A}$, the emptiness value, universality value of it, denoted by $e\_val(\mathcal{A})$,
$u\_val(\mathcal{A})$ respectively, are defined as:

$e\_val(\mathcal{A})=\bigvee\{L_{\omega}(\mathcal{A})(w)|w\in \Sigma^{\omega}\}$,

$u\_val(\mathcal{A})=\bigwedge\{L_{\omega}(\mathcal{A})(w)|w\in \Sigma^{\omega}\}$.

The emptiness-value (universality-value) problem for $\mathcal{A}$ is to decide whether $e\_val(\mathcal{A})\sim l$ ($u\_val(\mathcal{A})\sim l$), where $\sim$ is an order relation of $\{<,\leq,=,\geq,>\}$ and $l$ is a value of $L$.

\begin{Th}
The emptiness-value problem and universality-value problem for $L$-fuzzy alternating $B\ddot{u}chi$ automata are decidable in exponential time and are PSPACE-complete.
\end{Th}

\p{From the fact that the emptiness-value problem for $L$-fuzzy nondeterministic $\mathrm{B\ddot{u}chi}$ automata is decidable in linear time, the languages recognized by $L$-fuzzy nondeterministic $\mathrm{B\ddot{u}chi}$ automata are NLOGSPACE, and the unavoidable exponential blow-up of states is involved in the translation from an $L$-fuzzy nondeterministic $\mathrm{B\ddot{u}chi}$ automaton to its equivalent $L$-fuzzy alternating $\mathrm{B\ddot{u}chi}$ automaton, we know that the emptiness-value problem for $L$-fuzzy alternating $\mathrm{B\ddot{u}chi}$ automata are decidable in exponential time and the languages recognized by them are PSPACE.

All that remains to be proven is that the PSPACE-hardness of emptiness-value problem. In fact, it is easy to be shown similarly to Proposition 21 in \cite{22}: we reduce the emptiness-value problem for alternating automata to one for $L$-fuzzy alternating automata, and moreover, reduce the latters to another one for $L$-fuzzy alternating $\mathrm{B\ddot{u}chi}$ automata. Since the emptiness-value problem for alternating automata is PSPACE-complete, then emptiness-value problem for $L$-fuzzy alternating $\mathrm{B\ddot{u}chi}$ automata is also PSPACE-complete.

Afterwards, we consider the universality-value problem. Because the universality-value problem is dual to the emptiness-value problem and the complementation construction for $L$-fuzzy alternating $\mathrm{B\ddot{u}chi}$ automata cannot cause the changes of the states, then we obtain the university-value problem for $L$-fuzzy alternating $\mathrm{B\ddot{u}chi}$ automata is decidable in exponential time and having PSPACE-complexity.}

Considering two $L$-fuzzy alternating $\mathrm{B\ddot{u}chi}$ automata $\mathcal{A}_{1}$ and $\mathcal{A}_{2}$ over an identical lattice, the implication value of $\mathcal{A}_{1}$ with respect to $\mathcal{A}_{2}$ is defined as:

$imp\_value(\mathcal{A}_{1},\mathcal{A}_{2})
=\bigwedge\limits_{w\in\Sigma^{\omega}}(c(L_{\omega}(\mathcal{A}_{1})(w))\vee L_{\omega}(\mathcal{A}_{2})(w))$.

In addition, for any two $L$-fuzzy alternating $\mathrm{B\ddot{u}chi}$ automata $\mathcal{A}_{1}$ and $\mathcal{A}_{2}$ and a value $l\in L$, the implication-value problem is to decide whether $imp\_value(\mathcal{A}_{1},\mathcal{A}_{2})\sim l$, where $l$ is an order relation of $\{<,\leq,=,\geq,>\}$.

\begin{Th}
The implication-value for $L$-fuzzy alternating $B\ddot{u}chi$ automata are decidable in exponential time and are PSPACE-complete.
\end{Th}

Note that $imp\_value(\mathcal{A}_{1},\mathcal{A}_{2})\sim l$ if and only if $e\_val(\mathcal{A}_{1}\wedge \overline{\mathcal{A}_{2}})\sim^{\prime} c(l)$, where $<^{\prime},\leq^{\prime},=^{\prime},\geq^{\prime},>^{\prime}$ are $>,\geq,=,\leq,<$ respectively. Moreover,
$\mathcal{A}_{1}\wedge \overline{\mathcal{A}_{2}}$ is an $L$-fuzzy alternating $\mathrm{B\ddot{u}chi}$ automata recognizing the meet of $\mathcal{A}_{1}$ and the complement of $\mathcal{A}_{2}$ (Theorem 4.1 and 4.2), and its size is linear in $\mathcal{A}_{1}$ and linear in $\mathcal{A}_{2}$. So, the conclusion can be obtained.

\section{Illustrative examples}
\label{6}

In this section, we will give three examples to illustrate how to put our theories into the specific calculations. The first one is to construct an equivalent $L$-fuzzy nondeterministic $\mathrm{B\ddot{u}chi}$ automaton for a given $L$-fuzzy alternating $\mathrm{B\ddot{u}chi}$ automaton.

\begin{Exam}
Let $\mathcal{A}=(Q,\Sigma,\delta,I,F)$ be an $L$-fuzzy alternating $B\ddot{u}chi$ automaton, where

$L=([0,1],\vee,\wedge,0,1)$;
$Q=\{q_{0},q_{1},q_{2}\}$;
$\Sigma=\{a,b\}$;

$I(q_{0})=0.6$,
$I(q_{1})=I(q_{2})=0$;
$F(q_{0})=0$,
$F(q_{1})=0.4$,
$F(q_{2})=0.8$;

$\delta(q_{0},a)=0.7\wedge q_{1}$,
$\delta(q_{0},b)=(0.5\wedge q_{2})\vee 0.3$,
$\delta(q_{1},a)=q_{1}\wedge q_{2}$,
$\delta(q_{1},b)=q_{2}$,
$\delta(q_{2},a)=false$,
$\delta(q_{2},b)=q_{2}$.

It's not very hard to see that there are four successful run trees of $\mathcal{A}$, and we use $r_{i}$ ($i=1,\cdots,4$) to denote them, of which
$r_{1}$ and $r_{2}$ are the successful runs on $w_{1}=aab^{\omega}$ and
$w_{2}=ab^{\omega}$ respectively;
$r_{3}$ and $r_{4}$ are successful ones on $w_{3}=b^{\omega}$, and simultaneously
$r_{4}$ is a successful one on each word $w_{4}\in b\Sigma^{\omega}-\{b^{\omega}\}$ (cf. Figure 2.), then we have:

$L_{\omega}(\mathcal{A})(w_{1})=I(r_{1}(\varepsilon))\wedge wt(r_{1})=0.6\wedge 0.7\wedge 0.8=0.6$;

$L_{\omega}(\mathcal{A})(w_{2})=I(r_{2}(\varepsilon))\wedge wt(r_{2})=0.6\wedge 0.7\wedge 0.8=0.6$.

$L_{\omega}(\mathcal{A})(w_{3})=(I(r_{3}(\varepsilon))\wedge wt(r_{3}))\vee(I(r_{4}(\varepsilon))\wedge wt(r_{4}))=(0.6\wedge 0.5\wedge 0.8)\vee 0.3=0.5$.

$L_{\omega}(\mathcal{A})(w_{4})=I(r_{4}(\varepsilon))\wedge wt(r_{4})=0.3$.

\begin{center}
  $\includegraphics[height=6cm,width=9cm]{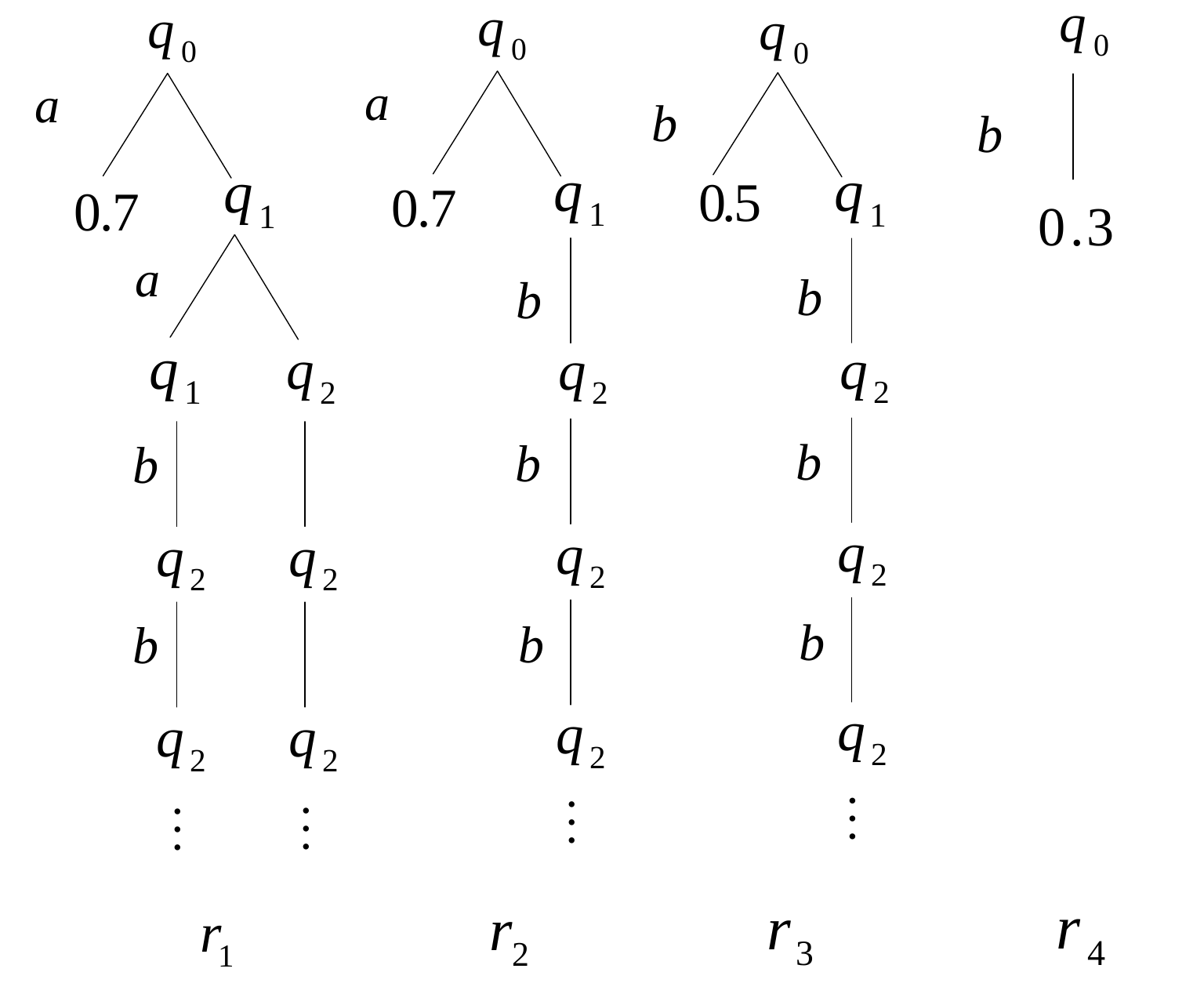}$
\end{center}
\begin{center}
  Figure 2: All successful runs of $\mathcal{A}$
\end{center}

\begin{center}
  $\includegraphics[height=6cm,width=9cm]{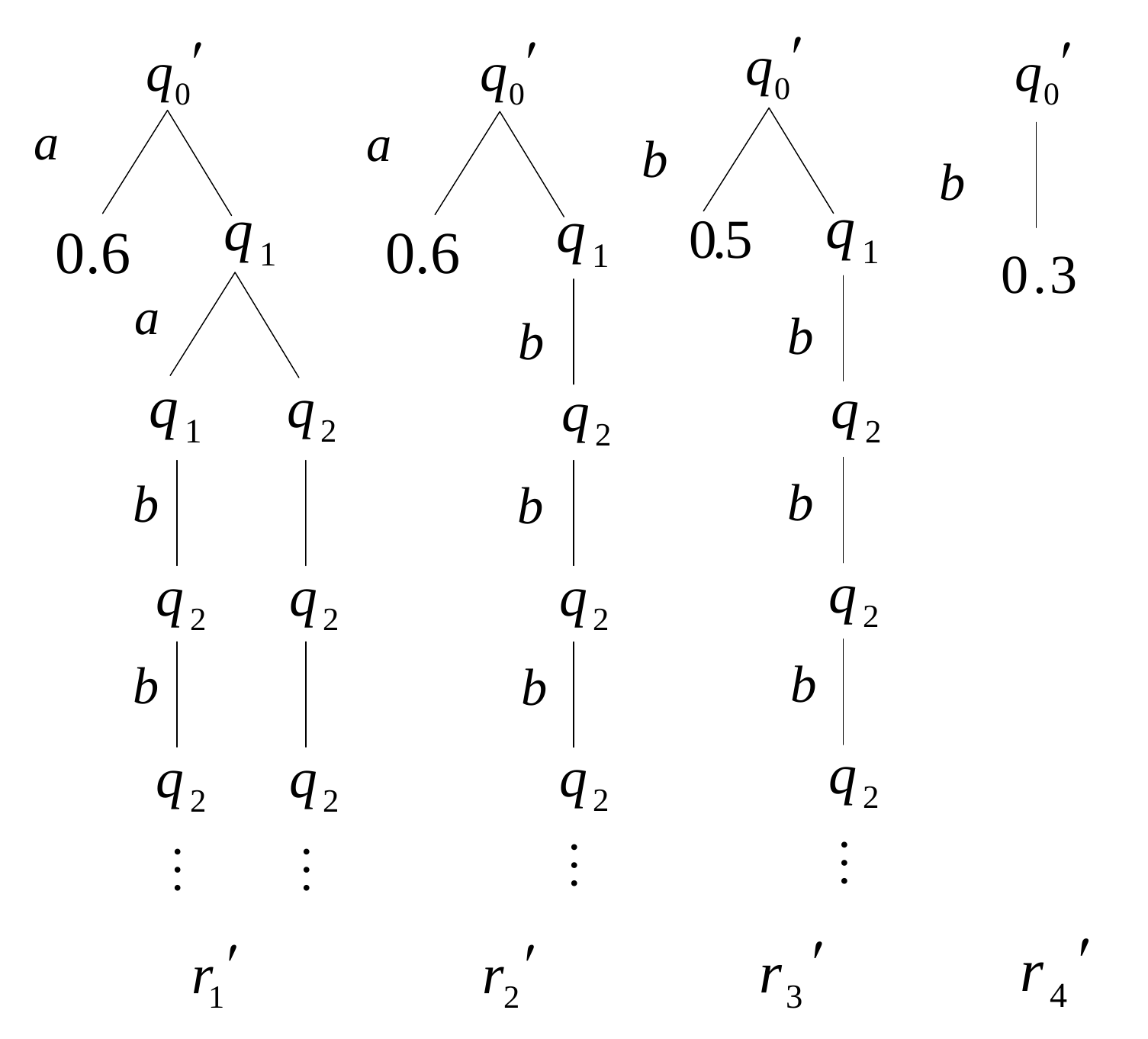}$
\end{center}
\begin{center}
  Figure 3: All successful runs of $\mathcal{A}^{\prime}$
\end{center}

According to Remark 3.4, we firstly construct its equivalent $L$-fuzzy alternating $B\ddot{u}chi$ automaton with a crisp initial state $\mathcal{A}^{\prime}=(Q^{\prime},\Sigma,\delta^{\prime},q_{0}^{\prime},F^{\prime})$:

$Q^{\prime}=Q\cup\{q_{0}^{\prime}\}$ ($q_{0}^{\prime}\notin Q$);

$F(q_{0}^{\prime})=F(q_{0})=0$,
$F(q_{1})=0.4$,
$F(q_{2})=0.8$;

$\delta(q_{0}^{\prime},a)=0.6\wedge q_{1}$,
$\delta(q_{0}^{\prime},b)=(0.5\wedge q_{2})\vee 0.3$,

$\delta(q_{0},a)=0.7\wedge q_{1}$,
$\delta(q_{0},b)=(0.5\wedge q_{2})\vee 0.3$,

$\delta(q_{1},a)=q_{1}\wedge q_{2}$,
$\delta(q_{1},b)=q_{2}$,

$\delta(q_{2},a)=false$,
$\delta(q_{2},b)=q_{2}$.

The corresponding successful runs are shown in Figure 3.

Because $|supp(F^{\prime})|=2$, then we construct an equivalent $L$-fuzzy alternating $B\ddot{u}chi$ automaton with crisp final states $\mathcal{A}^{\prime\prime}$ secondly, which is obtained by four
$L$-fuzzy nondeterministic $B\ddot{u}chi$ automata $\mathcal{A}_{i}^{\prime}$ ($i=1,\cdots,4$):

From Proposition 3.9, we know $ker(F)\cup 1(Q^{\prime})\cup 2(Q^{\prime})=\{\emptyset,\{q_{1}\},\{q_{2}\},\{q_{1},\\q_{2}\}\}$, then correspondingly, we construct $\mathcal{A}_{\emptyset}^{\prime}$,
$\mathcal{A}_{\{q_{1}\}}^{\prime}$,
$\mathcal{A}_{\{q_{2}\}}^{\prime}$,
$\mathcal{A}_{\{q_{1},q_{2}\}}^{\prime}$:

$\mathcal{A}_{\emptyset}^{\prime}=(Q^{\prime},\Sigma,\delta,I_{\emptyset},\emptyset)$,
where $q_{0}^{\prime}$ is the unique initial state with initial weight $I_{\emptyset}(q_{0}^{\prime})=1$. Since the final states set of $\mathcal{A}_{\emptyset}^{\prime}$ is empty, then we know every $r_{i}$ ($i=1,\cdots,4$) aren't successful in $\mathcal{A}_{\emptyset}^{\prime}$.

$\mathcal{A}_{\{q_{1}\}}^{\prime}=(Q^{\prime},\Sigma,\delta,I_{\{q_{1}\}},\{q_{1}\})$,
where the unique initial state is $q_{0}^{\prime}$ and $I_{\emptyset}(q_{0}^{\prime})=F^{\prime}(q_{1})=F(q_{1})$; Note that
$r_{i}$ ($i=1,\cdots,4$) aren't successful in $\mathcal{A}_{\{q_{1}\}}^{\prime}$ similarly, and thus, $weight_{\mathcal{A}_{\{q_{1}\}}^{\prime}}(r_{i})=0$ ($i=1,\cdots,4$).

$\mathcal{A}_{\{q_{2}\}}^{\prime}=(Q^{\prime},\Sigma,\delta,I_{\{q_{2}\}},\{q_{2}\})$,
where the initial weight $I_{\emptyset}(q_{0}^{\prime})$ of a unique initial state
$q_{0}^{\prime}$ is $F^{\prime}(q_{2})$ ($=F(q_{2})$), and therefore, we have $weight_{\mathcal{A}_{\{q_{2}\}}^{\prime}}(r_{i})=weight_{\mathcal{A}^{\prime}}(r_{i})$ for $i=1,\cdots,4$.

$\mathcal{A}_{\{q_{1},q_{2}\}}^{\prime}=(Q^{\prime},\Sigma,\delta,I_{\{q_{1},q_{2}\}},\{q_{1},q_{2}\})$,
where the unique non-zero initial weight of initial state is $I_{\emptyset}(q_{0}^{\prime})=F^{\prime}(q_{1})\wedge F^{\prime}(q_{2})=F(q_{1})\wedge F(q_{2})$, and $weight_{\mathcal{A}_{\{q_{2}\}}^{\prime}}(r_{i})$ is less than or equal to $ weight_{\mathcal{A}^{\prime}}(r_{i})$ ($i=1,\cdots,4$).

Renaming the states set of these four such that any two states sets are disjointed, we can obtain the new four automata, denoted by $\mathcal{A}_{i}^{\prime}$ ($i=1,\cdots,4$),
$\mathcal{A}_{i}^{\prime}=(Q_{i},\Sigma,\delta,I_{i},F_{i})$, where $Q_{i}=\{(q_{0}^{\prime})^{(i)},q_{0}^{(i)},q_{1}^{(i)},q_{2}^{(i)}\}$;

Their non-zero initial weights of initial states are:
$I_{1}((q_{0}^{\prime})^{(1)})=1$,
$I_{2}((q_{0}^{\prime})^{(2)})\\=F(q_{1})$,
$I_{3}((q_{0}^{\prime})^{(3)})=F(q_{2})$,
$I_{4}((q_{0}^{\prime})^{(4)})=F(q_{1})\wedge F(q_{2})$, and transition $\delta_{i}(q_{j}^{(i)},a)$ is obtained by instituting $q^{\prime}$ by $(q^{\prime})^{(i)}$ in original $\delta(q_{j},a)$, for any $q_{j}^{(i)}\in Q_{i},a\in \Sigma,j=1,\cdots,4$.
Their final states set are $F_{1}=\emptyset$,
$F_{2}=\{q_{1}^{(2)}\}$,
$F_{3}=\{q_{2}^{(3)}\}$,
$F_{4}=\{q_{1}^{(4)},q_{2}^{(4)}\}$ respectively.

\begin{center}
  $\includegraphics[height=12cm,width=10cm]{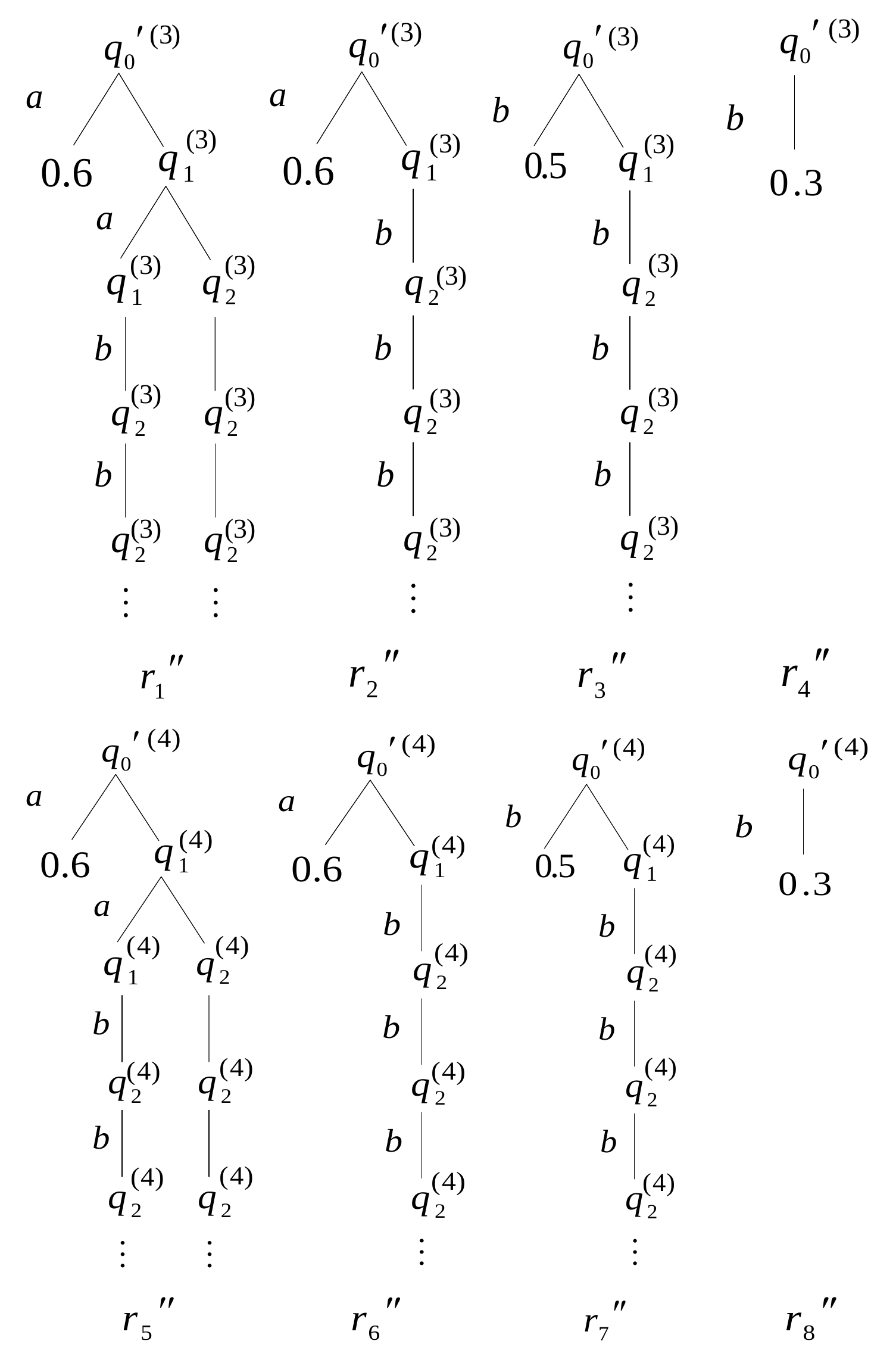}$
\end{center}
\begin{center}
  Figure 4: All successful runs of $\mathcal{A}^{\prime\prime}$
\end{center}

Put $\mathcal{A}^{\prime\prime}=(Q^{\prime\prime},\Sigma,\delta^{\prime\prime},I^{\prime\prime},F^{\prime\prime})$, where $Q^{\prime\prime}=\bigcup\limits_{i=1}^{4}Q_{i}$;
$I^{\prime\prime}(q)=I_{i}(q)$, if $q\in Q_{i}$;
$\delta^{\prime\prime}(q,a)=\delta_{i}(q,a)$, if $q\in Q_{i}$;
$F^{\prime\prime}=\{q_{1}^{(2)},q_{2}^{(3)},q_{1}^{(4)},q_{2}^{(4)}\}$. It's not very hard to see that there are eight successful runs of $\mathcal{A}^{\prime\prime}$ shown in Figure 4. For each $i\in\{1,\cdots,4\}$, $L_{\omega}(\mathcal{A}^{\prime\prime})(w_{i})=\bigvee\limits_{i=1}^{4}L_{\omega}(\mathcal{A}_{i}^{\prime})(w_{i})$ holds, and for other words, the weights of them are $0$.

At last, we apply Proposition 3.7 to obtain our desired equivalent $L$-fuzzy nondeterministic $B\ddot{u}chi$ automaton $\mathcal{A}_{n}$, where $\mathcal{A}_{n}=(Q_{n},\Sigma,\delta_{n},I_{n},F_{n})$ and

$Q_{n}=2^{Q^{\prime\prime}}\times 2^{Q^{\prime\prime}}$;

$I_{n}((\{(q_{0}^{\prime})^{(1)}\},\emptyset))=1$,
$I_{n}((\{(q_{0}^{\prime})^{(2)}\},\emptyset))=0.4$,
$I_{n}((\{(q_{0}^{\prime})^{(3)}\},\emptyset))=0.8$,
$I_{n}((\{(q_{0}^{\prime})^{(4)}\},\emptyset))=0.4$;
$F_{n}=\{\emptyset\}\times 2^{Q^{\prime\prime}}$;

$\delta_{n}((\{(q_{0}^{\prime})^{(3)}\},\emptyset),a,(\{q_{1}^{(3)}\},\{q_{1}^{(3)}\}))=0.6$,

$\delta_{n}((\{(q_{0}^{\prime})^{(3)}\},\emptyset),b,(\{q_{1}^{(3)}\},\{q_{1}^{(3)}\}))=0.5$,

$\delta_{n}((\{(q_{0}^{\prime})^{(4)}\},\emptyset),a,(\{q_{1}^{(4)}\},\emptyset))=0.6$,
$\delta_{n}((\{(q_{0}^{\prime})^{(4)}\},\emptyset),b,(\{q_{1}^{(4)}\},\emptyset))=0.5$,

$\delta_{n}((\{(q_{0}^{\prime})^{(i)}\},\emptyset),b,(\emptyset,\emptyset))=0.3$,

$\delta_{n}((\{q_{1}^{(3)}\},\{q_{1}^{(3)}\}),a,(\{q_{1}^{(3)},q_{2}^{(3)}\},\{q_{1}^{(3)}\}))=1$,

$\delta_{n}((\{q_{1}^{(3)}\},\{q_{1}^{(3)}\}),b,(\{q_{2}^{(3)}\},\emptyset))=1$,
$\delta_{n}((\{q_{1}^{(4)}\},\emptyset),a,(\{q_{1}^{(4)},q_{2}^{(4)}\},\emptyset))=1$,

$\delta_{n}((\{q_{1}^{(4)}\},\emptyset),b,(\{q_{2}^{(4)}\},\emptyset))=1$,
$\delta_{n}((\{q_{2}^{(i)}\},\emptyset),b,(\{q_{2}^{(i)}\},\emptyset))=1$,

$\delta_{n}((\{q_{1}^{(3)},q_{2}^{(3)}\},\{q_{1}^{(3)}\}),b,(\{q_{2}^{(3)}\},\emptyset))=1$,

$\delta_{n}((\{q_{1}^{(4)},q_{2}^{(4)}\},\emptyset),b,(\{q_{2}^{(4)}\},\emptyset))=1$

$\delta_{n}((\emptyset,\emptyset),a,(\emptyset,\emptyset))=1$,
$\delta_{n}((\emptyset,\emptyset),b,(\emptyset,\emptyset))=1$ (the $i$ occurring in above transitions merely could be $3$ or $4$, and the weight of transitions not mentioned are $0$).

Then the successful pathes of $\mathcal{A}_{n}$ are:

$P_{1}:(\{(q_{0}^{\prime})^{(3)}\},\emptyset)\stackrel{a/0.6}{\longrightarrow}(\{q_{1}^{(3)}\},\{q_{1}^{(3)}\})
\stackrel{a/1}{\longrightarrow}(\{q_{1}^{(3)},q_{2}^{(3)}\},\{q_{1}^{(3)}\})\stackrel{b/1}{\longrightarrow}
(\{q_{2}^{(3)}\},\emptyset)\\
\stackrel{b/1}{\longrightarrow}(\{q_{2}^{(3)}\},\emptyset) \cdots$,

$P_{2}:(\{(q_{0}^{\prime})^{(4)}\},\emptyset)\stackrel{a/0.6}{\longrightarrow}(\{q_{1}^{(4)}\},\emptyset)
\stackrel{a/1}{\longrightarrow}(\{q_{1}^{(4)},q_{2}^{(4)}\},\emptyset)\stackrel{b/1}{\longrightarrow}
(\{q_{2}^{(4)}\},\emptyset)
\stackrel{b/1}{\longrightarrow}(\{q_{2}^{(4)}\},\emptyset) \cdots$,

$P_{3}:(\{(q_{0}^{\prime})^{(3)}\},\emptyset)\stackrel{a/0.6}{\longrightarrow}(\{q_{1}^{(3)}\},\{q_{1}^{(3)}\})
\stackrel{b/1}{\longrightarrow}(\{q_{2}^{(3)}\},\emptyset)\stackrel{b/1}{\longrightarrow}(\{q_{2}^{(3)}\},\emptyset)
\cdots$,

$P_{4}:(\{(q_{0}^{\prime})^{(4)}\},\emptyset)\stackrel{a/0.6}{\longrightarrow}(\{q_{1}^{(4)}\},\emptyset)
\stackrel{b/1}{\longrightarrow}(\{q_{2}^{(4)}\},\emptyset)\stackrel{b/1}{\longrightarrow}(\{q_{2}^{(4)}\},\emptyset)
\cdots$,

$P_{5}:(\{(q_{0}^{\prime})^{(3)}\},\emptyset)\stackrel{b/0.5}{\longrightarrow}(\{q_{1}^{(3)},\{q_{1}^{(3)})
\stackrel{b/1}{\longrightarrow}(\{q_{2}^{(3)}\},\emptyset)\stackrel{b/1}{\longrightarrow}(\{q_{2}^{(3)}\},\emptyset)
\cdots$,

$P_{6}:(\{(q_{0}^{\prime})^{(4)}\},\emptyset)\stackrel{b/0.5}{\longrightarrow}(\{q_{1}^{(4)}\},\emptyset)
\stackrel{b/1}{\longrightarrow}(\{q_{2}^{(4)}\},\emptyset)\stackrel{b/1}{\longrightarrow}(\{q_{2}^{(4)}\},\emptyset)
\cdots$,

$P_{j\geq 7}(i=3\ or \ 4):(\{(q_{0}^{\prime})^{(i)}\},\emptyset)\stackrel{b/0.3}{\longrightarrow}(\emptyset,\emptyset)
\stackrel{a,b/1}{\longrightarrow}(\emptyset,\emptyset)\stackrel{a,b/1}{\longrightarrow}(\emptyset,\emptyset)\cdots$,

Therefore, we have
$L_{\omega}(\mathcal{A}_{n})(w_{1})=0.6$,
$L_{\omega}(\mathcal{A}_{n})(w_{2})=0.6$,
$L_{\omega}(\mathcal{A}_{n})(w_{3})=0.5$,
$L_{\omega}(\mathcal{A}_{n})(w_{4})=0.3$, and the other weights are $0$, which shows that $\mathcal{A}_{n}$ is an $L$-fuzzy nondeterministic $B\ddot{u}chi$ automaton equivalent to $\mathcal{A}$, as required.

\end{Exam}

The next example can verify the correctness about the closure property about complement of $L$-fuzzy alternating $\mathrm{B\ddot{u}chi}$ automata by taking dual operation and changing the final weights to their complements.

\begin{Exam}
We begin with the $\mathcal{A}^{\prime}$ in the previous example. It is easy to see its dual $\overline{\mathcal{A}^{\prime}}$ is $(Q^{\prime},\Sigma,\overline{\delta^{\prime}},q_{0}^{\prime},(F^{\prime})^{c})$, where

$Q^{\prime}=Q\cup\{q_{0}^{\prime}\}$ ($q_{0}^{\prime}\notin Q$);
$c(a)=1-a$;$F^{c}(q_{0}^{\prime})=1$,

$F^{c}(q_{0})=1$,
$F^{c}(q_{1})=0.6$,
$F^{c}(q_{2})=0.2$;

$\delta(q_{0}^{\prime},a)=0.4\vee q_{1}$,
$\delta(q_{0}^{\prime},b)=(0.5\vee q_{2})\wedge 0.3$,
$\delta(q_{0},a)=0.3\vee q_{1}$,
$\delta(q_{0},b)=(0.5\vee q_{2})\wedge 0.7$,
$\delta(q_{1},a)=q_{1}\vee q_{2}$,
$\delta(q_{1},b)=q_{2}$,
$\delta(q_{2},a)=true$,
$\delta(q_{2},b)=q_{2}$.

\begin{center}
  $\includegraphics[height=11cm,width=10cm]{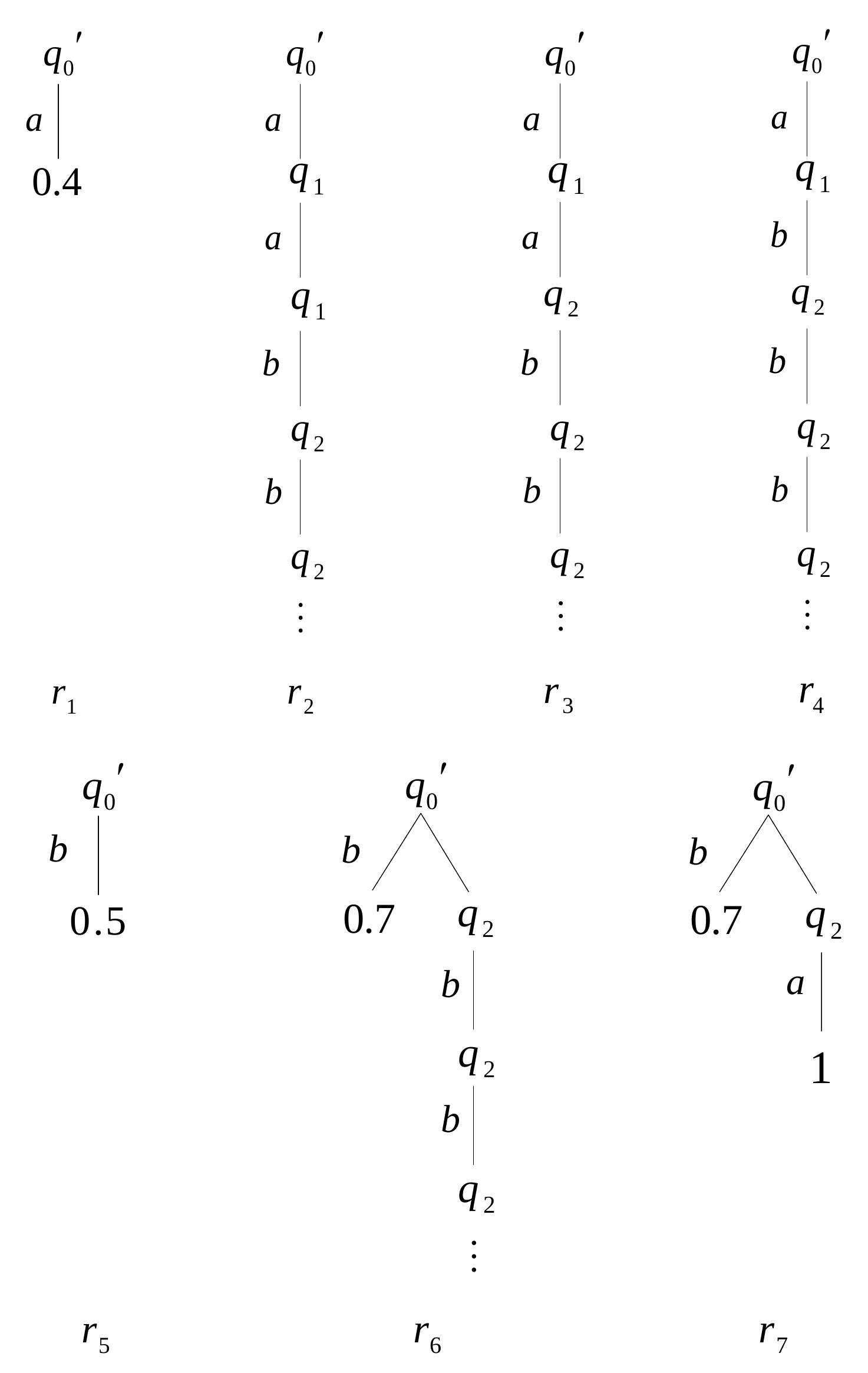}$
\end{center}
\begin{center}
  Figure 5: All successful runs of $\overline{\mathcal{A}^{\prime}}$
\end{center}

There are seven successful runs of $\overline{\mathcal{A}^{\prime}}$, denoted by $r_{i}(1=1,\cdots,7)$ (cf. Figure 5.), of which $r_{1},r_{2},r_{3}$ are successful ones on $aab^{\omega}$;
$r_{1}$ and $r_{4}$ are successful ones on $ab^{\omega}$;
$r_{5},r_{6}$ are successful on $b^{\omega}$; Simultaneously,
$r_{5}$ and $r_{7}$ are successful one on each word $w\in b\Sigma^{\ast}-\{b^{\omega}\}$, and therefore, we have:

$L_{\omega}(\overline{\mathcal{A}^{\prime}})(aab^{\omega})=\bigvee\limits_{i=1}^{3}wt(r_{i})=0.4$,
$L_{\omega}(\overline{\mathcal{A}^{\prime}})(ab^{\omega})=wt(r_{1})\vee wt(r_{5})=0.4$,

$L_{\omega}(\overline{\mathcal{A}^{\prime}})(b^{\omega})=wt(r_{5})\vee wt(r_{6})=0.5$,
$L_{\omega}(\overline{\mathcal{A}^{\prime}})(w)=wt(r_{5})\vee wt(r_{7})=0.7$.

All that remains to be proven is that for any $w^{\prime}\in a\Sigma^{\omega}-\{aab^{\omega}\}-\{ab^{\omega}\}$,
$L_{\omega}(\overline{\mathcal{A}^{\prime}})(w^{\prime})=1$. In fact, there are two possibilities:

If $w^{\prime}\in aab\Sigma^{\omega}-\{aabb^{\omega}\}$, i.e., from the third input symbol,
$b$, there is at least a symbol,
$a$, appearing in $w^{\prime}$, then there is a successful (finite) run hit the true transition, therefore, the largest weight of successful run on $w^{\prime}$ is $1$, and thus,
$L_{\omega}(\overline{\mathcal{A}^{\prime}})(w^{\prime})=1$.

If $w^{\prime}\in ab\Sigma^{\omega}-\{ab^{\omega}\}$, i.e., from the second input symbol,
$b$, there is at least a symbol,
$a$, appearing in $w^{\prime}$, then there is a successful (finite) run hit the true transition similarly, so
$L_{\omega}(\overline{\mathcal{A}^{\prime}})(w^{\prime})=1$ holds.

\end{Exam}

The last example taken by us is to present how to transform an $L$-fuzzy alternating co-$\mathrm{B\ddot{u}chi}$ automaton to its equivalent $\mathrm{B\ddot{u}chi}$ one. The several identical procedures with respect to Example 6.1 below will be omitted.

\begin{Exam}
Let $\mathcal{A}=(Q,\Sigma,\delta,q_{0},F)$ be such a co-$B\ddot{u}chi$ one, where

$L=([0,1],\vee,\wedge,0,1)$;
$c(a)=1-a$;
$Q=\{q_{0},q_{1}\}$;
$\Sigma=\{a,b\}$;

$F(q_{0})=0.4$,
$F(q_{1})=0.8$;
$\delta(q_{0},a)=0.7\wedge q_{1}$,
$\delta(q_{0},b)=(0.5\wedge q_{1})\vee 0.3$,

$\delta(q_{1},a)=q_{0}\wedge q_{1}$,
$\delta(q_{1},b)=false$.

\begin{center}
  $\includegraphics[height=5.5cm,width=9cm]{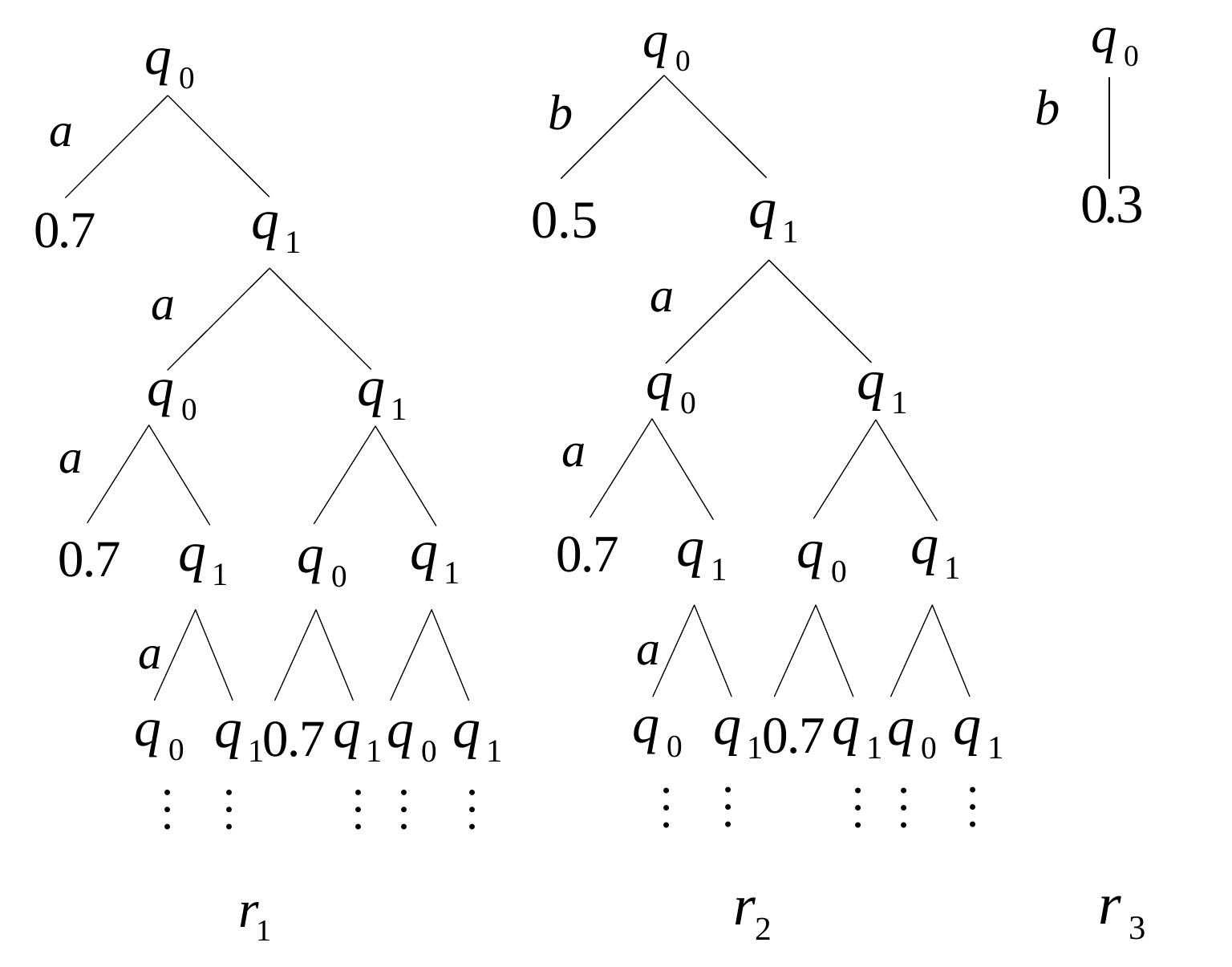}$
\end{center}
\begin{center}
  Figure 6: All successful runs of $\mathcal{A}$
\end{center}

It is easy to see that there are three runs of $\mathcal{A}$, we denote them by $r_{i}$ ($i=1,2,3$) (cf. Figure 6.) and the corresponding weights of words are:

$L_{\omega}(\mathcal{A})(a^{\omega})=wt(r_{1})=0.6\wedge 0.7\wedge 0.4=0.4$,

$L_{\omega}(\mathcal{A})(ba^{\omega})=wt(r_{1})\vee wt(r_{3})=(0.5\wedge 0.7\wedge 0.4)\vee 0.3=0.4$,

$L_{\omega}(\mathcal{A})(w)=0.3$, for any $w\in b\Sigma^{\omega}-\{ba^{\omega}\}$.

Firstly we turn it to another $L$-fuzzy alternating co-$B\ddot{u}chi$ automaton with only crisp final states by Proposition 4.5.
The first step of such process is to construct the dual of $\mathcal{A}$,
an $L$-fuzzy alternating $B\ddot{u}chi$ automaton $\overline{\mathcal{A}}=(Q,\Sigma,\overline{\delta},q_{0},F^{c})$, where

$F^{c}(q_{0})=0.6$,
$F^{c}(q_{1})=0.2$;

$\delta(q_{0},a)=0.3\vee q_{1}$,
$\delta(q_{0},b)=0.5\vee (0.7\wedge q_{1})$,

$\delta(q_{1},a)=q_{0}\vee q_{1}$,
$\delta(q_{1},b)=true$.

Secondly, constructing an equivalent $L$-fuzzy alternating $B\ddot{u}chi$ automaton $\mathcal{B}$ with crisp final states, similarly to Example 6.1, and the result automaton is:

$\mathcal{B}=(\widehat{Q},\Sigma,\overline{\delta}^{\prime},I^{\prime},\{q_{0}^{(2)},q_{1}^{(3)},q_{0}^{(4)},q_{1}^{(4)}\})$, where

$\widehat{Q}=\{q_{i}^{(j)}|i=0,1;j=1,\cdots,4\}$;
$I^{\prime}(q_{0}^{(1)})=1$,
$I^{\prime}(q_{0}^{(2)})=F^{c}(q_{0})=0.6$,

$I^{\prime}(q_{0}^{(3)})=F^{c}(q_{1})=0.2$,
$I^{\prime}(q_{0}^{(4)})=F^{c}(q_{0})\wedge F^{c}(q_{1})=0.2$;

$\overline{\delta}^{\prime}(q_{0}^{(i)},a)=0.3\vee q_{1}^{(i)}$,
$\overline{\delta}^{\prime}(q_{0}^{(i)},b)=0.5\vee (0.7\wedge q_{1}^{(i)})$,

$\overline{\delta}^{\prime}(q_{1}^{(i)},a)=q_{0}^{(i)}\vee q_{1}^{(i)}$,
$\overline{\delta}^{\prime}(q_{1}^{(i)},b)=true$.

Afterwards, we construct $\mathcal{B}^{\prime}$, an $L$-fuzzy alternating $B\ddot{u}chi$ automaton with a crisp initial state and crisp final states equivalent to $\mathcal{B}$ by adding an extra state and some transitions:

$\mathcal{B}^{\prime}=(Q^{\prime},\Sigma,\delta^{\prime},q_{0}^{\prime},\{q_{0}^{(2)},q_{1}^{(3)},q_{0}^{(4)},q_{1}^{(4)}\})$, where $q_{0}^{\prime}\notin Q^{\prime}$, and

\begin{eqnarray*}
\ \ \ \ \delta^{\prime}(q_{0}^{\prime},a)
&=&\bigvee\limits_{1=1}^{4}I^{\prime}(q_{0}^{(i)})\wedge \overline{\delta}^{\prime}(q_{0}^{(i)},a)\\
&=&0.3\vee q_{1}^{(1)}\vee (0.6\wedge q_{1}^{(2)})\vee 0.2\vee (0.2\wedge q_{1}^{(3)})\vee (0.2\wedge q_{1}^{(4)})\\
&\equiv&0.3\vee q_{1}^{(1)}\vee (0.6\wedge q_{1}^{(2)}),\\
\ \ \ \ \delta^{\prime}(q_{0}^{\prime},b)
&=&\bigvee\limits_{1=1}^{4}I^{\prime}(q_{0}^{(i)})\wedge \overline{\delta}^{\prime}(q_{0}^{(i)},b)\\
&=&0.5\vee (0.7\wedge q_{1}^{(1)})\vee (0.6\wedge q_{1}^{(2)})\vee 0.2\vee (0.2\wedge q_{1}^{(3)})\vee (0.2\wedge q_{1}^{(4)})\\
&\equiv&0.5\vee (0.7\wedge q_{1}^{(1)})\vee (0.6\wedge q_{1}^{(2)}),
\end{eqnarray*}

$\delta^{\prime}(q_{0}^{(i)},a)=0.3\vee q_{1}^{(i)}$,
$\delta^{\prime}(q_{0}^{(i)},b)=0.5\vee (0.7\wedge q_{1}^{(i)})$,

$\delta^{\prime}(q_{1}^{(i)},a)=q_{0}^{(i)}\vee q_{1}^{(i)}$,
$\delta^{\prime}(q_{1}^{(i)},b)=true$.

Further on, constructing the dual of $\mathcal{B}^{\prime}$. We can see that only $r_{1}^{\prime},r_{3}^{\prime},r_{4}^{\prime}$ are successful (cf. Figure 7.) and the corresponding weights of the languages are:

$L_{\omega}(\mathcal{B}^{\prime})(a^{\omega})=wt(r_{1}^{\prime})=0.4\wedge 0.7=0.4$,

$L_{\omega}(\mathcal{B}^{\prime})(ba^{\omega})=wt(r_{3}^{\prime})\vee wt(r_{4}^{\prime})=0.3\vee (0.4\wedge 0.7)=0.4$,

$L_{\omega}(\mathcal{B}^{\prime})(w)=0.3$, for any $w\in b\Sigma^{\omega}-\{ba^{\omega}\}$.

The last procedure is to build $\mathcal{B}^{\prime\prime}$, our desired (weak) $L$-fuzzy alternating $B\ddot{u}chi$ automaton, which is equivalent to the original $\mathcal{A}$:

$\mathcal{B}^{\prime\prime}=(Q^{\prime}\times [18],\Sigma,\delta^{\prime\prime},(q_{0}^{\prime},18),Q^{\prime}\times [18]^{odd})$ ($18=|Q^{\prime}|$), where (the others transitions not mentioned are false)

$\delta((q_{0}^{\prime},l),a)=(\bigvee\limits_{i\leq l}0.4\wedge (q_{1}^{(1)},i))\vee (\bigvee\limits_{i,j\leq l}0.7\wedge (q_{1}^{(1)},i)\wedge(q_{1}^{(2)},j))$,

$\delta((q_{0}^{\prime},l),b)=0.3\vee(\bigvee\limits_{i\leq l}0.4\wedge (q_{1}^{(1)},i))\vee (\bigvee\limits_{i,j\leq l}0.5\wedge (q_{1}^{(1)},i)\wedge(q_{1}^{(2)},j))$,

$\delta^{\prime\prime}((q_{0}^{(1)},l),a)=\bigvee\limits_{i\leq l}0.7\wedge (q_{1}^{(1)},i)$,

$\delta^{\prime\prime}((q_{0}^{(1)},l),b)=0.3\vee (\bigvee\limits_{i\leq l}0.5\wedge (q_{1}^{(1)},i))$,

$\delta^{\prime\prime}((q_{0}^{(2)},2l^{\prime}),a)=\bigvee\limits_{i\leq 2l^{\prime}}0.7\wedge (q_{1}^{(2)},i)$,

$\delta^{\prime\prime}((q_{0}^{(2)},2l^{\prime}),b)=0.3\vee (\bigvee\limits_{i\leq 2l^{\prime}}0.5\wedge (q_{1}^{(2)},i))$,

$\delta^{\prime\prime}((q_{1}^{(1)},l),a)=\bigvee\limits_{i,j\leq l}(q_{0}^{(1)},i)\wedge (q_{1}^{(1)},j)$,

$\delta^{\prime\prime}((q_{1}^{(2)},2l^{\prime}),a)=\bigvee\limits_{i,j\leq l}(q_{0}^{(2)},i)\wedge (q_{1}^{(2)},j)$,

$\delta^{\prime\prime}((q_{0}^{(3)},l),a)=\bigvee\limits_{i\leq l}0.7\wedge (q_{1}^{(3)},i)$,

$\delta^{\prime\prime}((q_{0}^{(3)},l),b)=0.3\vee (\bigvee\limits_{i\leq l}0.5\wedge (q_{1}^{(3)},i))$,

$\delta^{\prime\prime}((q_{1}^{(3)},2l^{\prime}),a)=\bigvee\limits_{i,j\leq 2l^{\prime}}(q_{0}^{(1)},i)\wedge (q_{1}^{(1)},j)$,

$\delta^{\prime\prime}((q_{0}^{(4)},2l^{\prime}),a)=\bigvee\limits_{i\leq 2l^{\prime}}0.7\wedge (q_{1}^{(4)},i)$,

$\delta^{\prime\prime}((q_{0}^{(4)},2l^{\prime}),b)=0.3\vee (\bigvee\limits_{i\leq 2l^{\prime}}0.5\wedge (q_{1}^{(4)},i))$,

$\delta^{\prime\prime}((q_{1}^{(4)},2l^{\prime}),a)=\bigvee\limits_{i,j\leq 2l^{\prime}}(q_{0}^{(4)},i)\wedge (q_{1}^{(4)},j)$ ($l,2l^{\prime} \ are \ the \ numbers \ less \ than\\ \ or \ equal \ to \ 18$).

The successful runs of $\mathcal{B}^{\prime\prime}$ are not only the following three, but their projections on $Q^{\prime}$ correspond to one of the projection of the three on $Q^{\prime}$ respectively, so their weights cannot make the whole languages of $\mathcal{B}^{\prime\prime}$ to be larger, then only considering following ones is enough (cf. Figure 8.). And, it is easy to examine that the $\mathcal{B}^{\prime\prime}$ is equivalent to the starting automaton $\mathcal{A}$.

\begin{center}
  $\includegraphics[height=14cm,width=12cm]{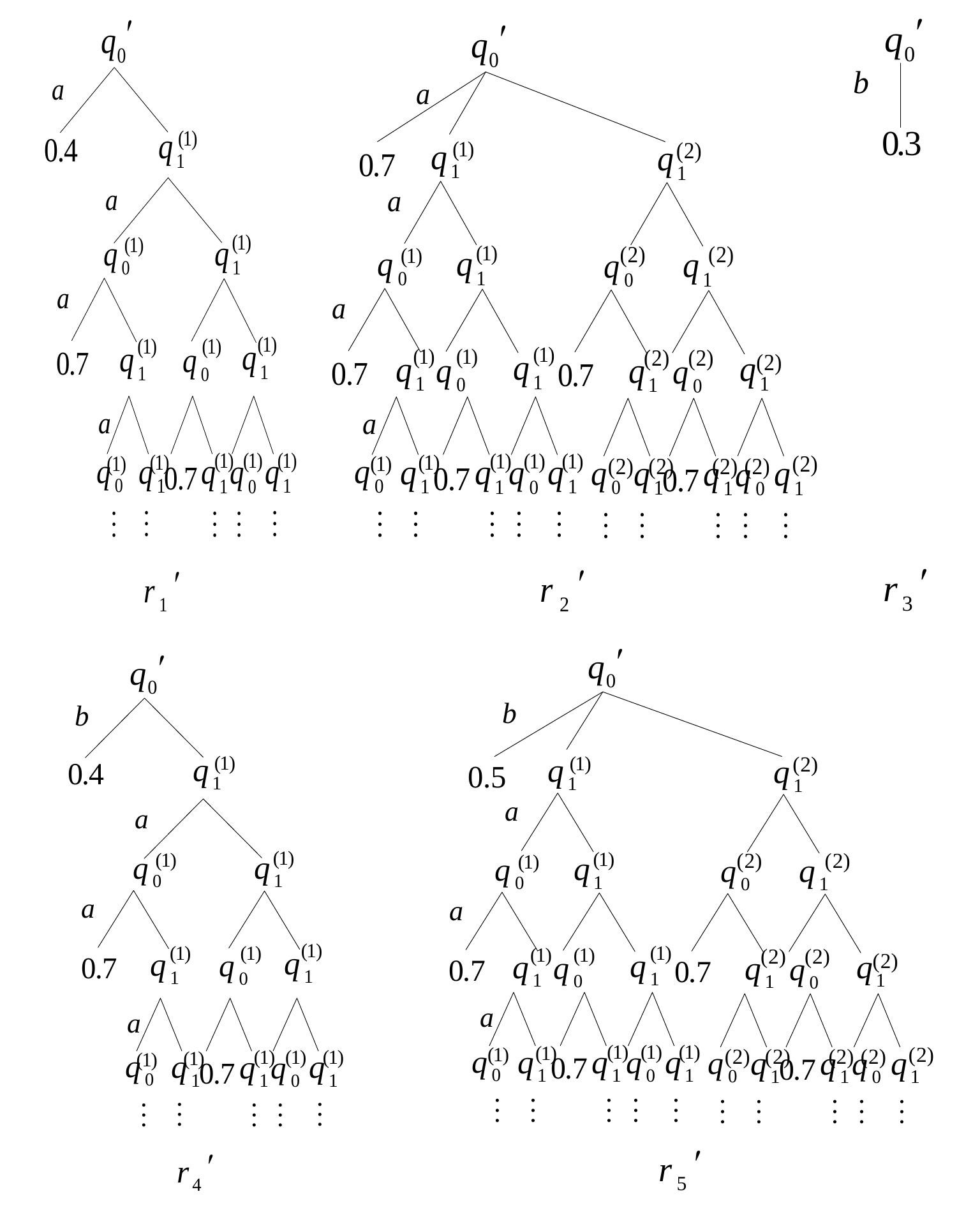}$
\end{center}
\begin{center}
  Figure 7: All successful runs of $\mathcal{B}^{\prime}$
\end{center}

\begin{center}
  $\includegraphics[height=15cm,width=12cm]{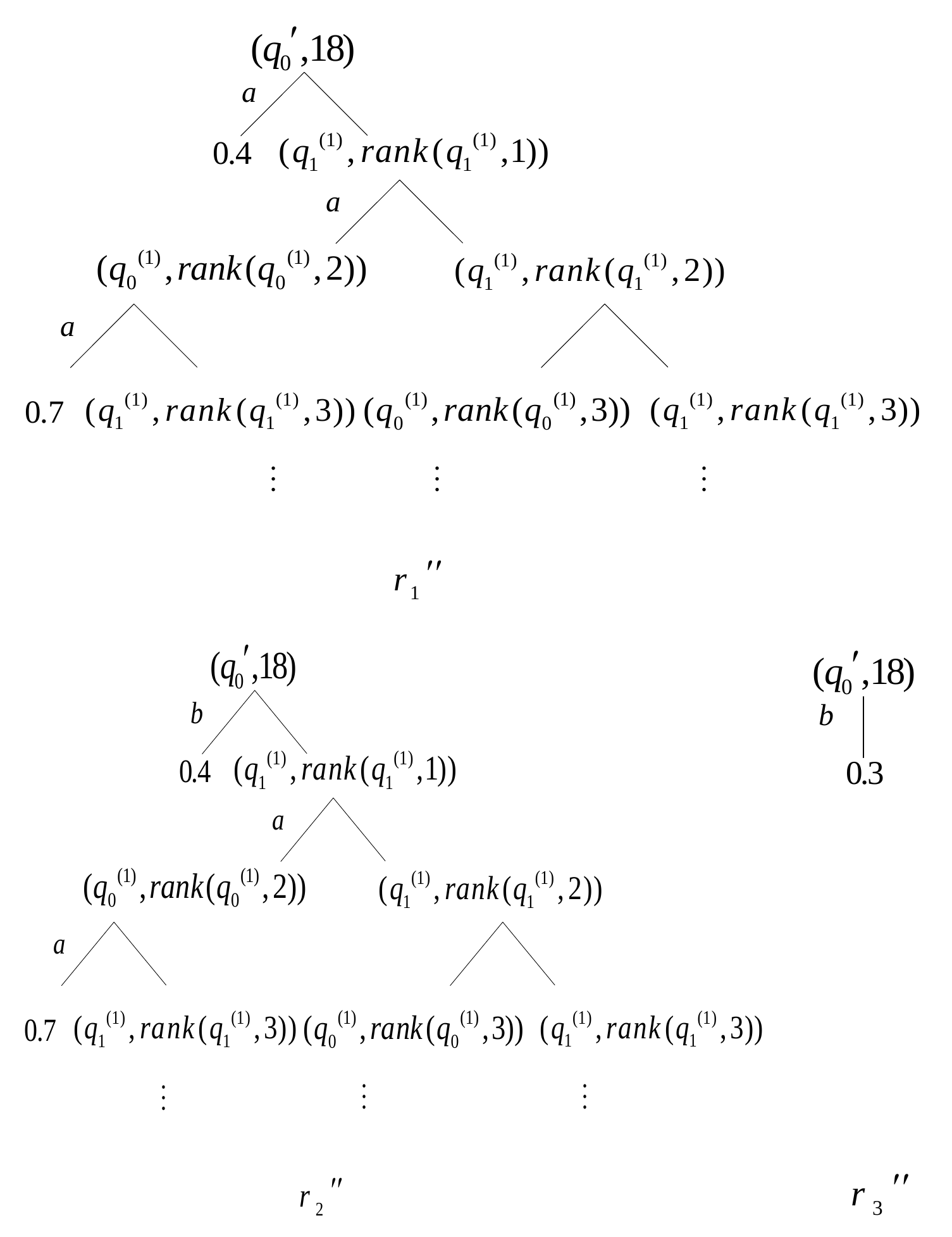}$
\end{center}
\begin{center}
  Figure 8: Some (not all) successful runs of $\mathcal{B}^{\prime\prime}$
\end{center}

\end{Exam}

\section{Conclusions}
\label{6}
The closure properties of $L$-fuzzy alternating $\mathrm{B\ddot{u}chi}$ automata and the equivalence relationship between them and $L$-fuzzy nondeterministic ones were already studied in our paper. We gave a direct construction to illustrate the $L$-fuzzy $\omega$-regularity of the languages recognized by $L$-fuzzy alternating co-$\mathrm{B\ddot{u}chi}$ automata without the related knowledge about $L$-fuzzy nondeterministic $\mathrm{B\ddot{u}chi}$ automata. In addition, the discussion about decision problems for $L$-fuzzy alternating $\mathrm{B\ddot{u}chi}$ automata and some illustrative examples were given in our paper. Using above preparations, we can study the properties about fuzzy temporal logic in model checking in the future, such as building a fuzzy alternating $\mathrm{B\ddot{u}chi}$ automaton for a given fuzzy LTL formula (\cite{15,16}) satisfying the languages of the automaton is exactly the fuzzy set of computations satisfying the formula.
\vspace{0.5cm}\\
\textbf{References}





\bibliographystyle{model1-num-names}



\end{document}